# Stretching and breaking of chemical bonds, correlation of electrons, and radical properties of covalent species

Elena Sheka

*Peoples' Friendship University of Russia,*
*117198 Moscow, Russia*
sheka@icp.ac.ru

Chemical bonds are considered in light of correlation of valence electrons that is strengthened when the bond is dissociated. In the framework of the unrestricted Hartree-Fock single-reference version of the configuration interaction theory, effectively unpaired electrons lay the foundation of the electron correlation measure in terms of total number of the electrons $N_D$ (molecular chemical susceptibility). $N_D(R)$ graphs and their singularities with respect to the interatomic distance $R$ allow presenting a quantitative description of stretching and breaking of chemical bonds. The approach validity is demonstrated on a large number of bonds of different order and chemical composition.

**Keywords**: chemical bond; covalent bond; chemical bond breaking; chemical bond stretching; electron correlation; effectively unpaired electrons; molecular chemical susceptibility

## 1. Introduction

The author, once involved in the computational chemistry for last two decades, has been dealing with chemical bonds as the main elements of structural chemistry that lay the foundation of any input-and/or-output structural model of the computations. The experience gained over these years sais that the presentation on standard fixed chemical bonds is not generally true since in many cases both contracted (less often) and stretched (a lot more) bonds are characteristic for equilibrium structures of computational objects. The finding raises a set of simple questions:
- What is a stretched/contracted chemical bond?
- What are reasons for the bond stretching/contraction?
- How much can be the bond stretched?
- At which length is the bond broken?

Obviously, all these questions are addressed to the very essence of chemical bonding within which should one look for answers.

A chemical bond was introduced in chemistry more then two centuries ago as the main concept to configure the attraction between atoms that provides the formation of a chemical substance. Since atom's electrons and nuclei are the main participants of the actions, the concept content in each historical period was a precise replica of the understanding of the electronic essence of the matters around us achieved by that time. The concept has now become one of the most fundamental operational aspects of modern chemistry. Covalent bonds, ionic bonds, metallic bonds, hydrogen bonds, and many others are common grounds of the modern chemical language. At the same time, each of them reflects a certain facet of the overall electronic theory of the matter.

Theoretically, the bond concept has come a long way of development alongside with the electron theory of chemical matter, and its development is still ongoing. Particular epochs are associated with the valence bond theory [1], molecular orbital theory [2], and density functional theory [3]. These theoretical approaches have laid the foundation of quantum chemistry aimed at obtaining equilibrium multi-atomic configurations. However, a direct solution of Schroedinger's equation does not point to the bond within a particular pair of atoms. Computationally, the bond justification consists in finding bond critical points related to the electron density distribution in the frame of either the atom-in-molecules theory [4] or some of its developments (see Ref. [5] and references therein). Empirically, in the majority of cases, the bond between two atoms is justified by comparing the interatomic distance with one of standard bond lengths accumulated on the basis of numerous structural data. In view of this interrelation, on practice, the chemical bond is mainly associated with this structural identificator with respect to which one can speak about 'bond forming', 'bond stretching', or 'bond breaking'. To the most extent, these 'bond transformations' are hot discussed with respect to covalent bonds.

Speaking about the length of a covalent bond, one usually addresses the data tabulated in numerous tables and presented in numerous handbooks (see, for example, Refs. [6, 7]). As seen from the data, bond lengths for the same pair of atoms in various molecules are rather consistent which makes it possible to speak about standard values related to particular pairs of atoms. Thus, a standard length of 1.09Å is attributed to the C-H pair while the lengths of 1.54, 1.34, and 1.20Å are related to single, double and triple C-C bond, respectively. Complicated as a whole, the set of the available data on bond lengths and bond energies provides a comprehensive view on the equilibrium state of molecules and solids. On the background of this self consistency, the detection of extremely long bonds, such as single C-C bonds of 1.647 Å, 1.659 Å, and 1.704 Å instead of 1.54 Å [8] and C-O bonds of 1.54 Å [9] and 1.622 Å [10] instead of 1.43 Å not only looks as a chemical curiosity but raises the question of the limits of covalent bonding. Two other questions are closely related to the latter: 1) to which extent a chemical bond can be stretched and 2) on which length its rupture occurs. Empirically, one can find subjectively made estimations of critical values of a possible elongation of bonds that widely varied. Thus, the width of the region of admissible values of bond's lengths significantly varied in different computer programs aimed at molecular imaging. As for a bond rupture, this problem is the most uncertain and the rupture is considered as a final result of a continuous stretching only.

The problem of theoretical justification of the chemical bond stretching and breaking concerns the criteria according to which the considered bond is still alive or ceases to exist. Until now, two approaches can be mentioned. The first one, based on the atom-in-molecules theory [4], concerns the bond critical point within the electron density distribution over an atomic composition, evidence of which is considered as a proof of the bond existence. However, as shown recently [11], the criterion, computationally realized, is not reliable in the case of weak coupling due to which it cannot be used to fix the bond breaking. The other approach overcomes the difficulty addressing directly to the correlation of electrons involved in the bond [12]. The concept of the entanglement among any pair of orbitals lays the approach foundation. In the framework of the information quantum theory, two entanglement measures, namely, the single orbital entropy $s(1)_i$ and the mutual information $I_{tot}$ are suggested to quantitatively describe the electron correlation while the relevant derivatives $\partial s(1)_i / \partial r_{AB} \to 0$ as well as $\partial I_{tot} / \partial r_{AB} \to 0$ may serve as an indication of either bond-forming or bond-breaking when the interatomic distance achieves $r_{AB}$. The approach has been recently applied for a thorough analysis of chemical bonds in $N_2$, $F_2$, and CsH molecules [13]. Calculations were carried out at the present level of the multireference configuration interaction (CI) theory. The entanglement measures were determined from wave functions optimized by the density matrix renormalization group (DMRG) while the complete active space self-consistent field (CASSCF) approach was used to configure the orbital basis in terms of natural orbitals. The obtained results showed that electron correlation is indeed the main determinant of stretching and breaking of chemical bonds and the quantitative measure of the correlation may serve as criteria for the fixation of the above processes. However, the computational procedure is time consuming, which postpones approaches' application to more complex molecules for the long term.

A quantitative description of electron correlation can be obtained not only in the frame work of the multireference CI theory, but exploiting particular properties of single-determinant solutions. The unrestricted Hartree-Fock (UHF) scheme is the best for the purpose. As shown in [14], three criteria can characterize the electron correlation: (i) misalignment of total energy; (ii) the appearance of effectively unpaired electrons; (iii) non-zero squared spin value for singlet molecules. The first criterion is characteristic for the CI theory of any level. Two other, once interconnected, are the consequence of the spin-mixed character of the UHF electronic state while opening a possibility of detailed description of the electron correlation with good accuracy [15]. Moreover, the HF level of the theory is quite sufficient for understanding the basic aspects of bonding [16]. These circumstances lead to new insights into the intrinsic features of chemical bonds from the viewpoint of electron correlation. The current chapter suggests the first realization of the UHF theory ability to establish criteria on forming, stretching, and breaking of chemical bonds. The disclosed trends are based on results obtained in the course of extended computational experiment that covered complete sets of chemical bonds $X \leftrightarrow X$, involving one-, two-, and three-electron ones, formed by atoms of group 14 (X= C, Si, Ge, and Sn) as well as one-electron single bonds $X - Y$ (Y= H, O). Additionally, the bond dissociation of $N_2$ and $F_2$ molecules were considered to have a possibility of comparing results of the multireference and single-reference CI theory. A particular attention is given to bonds' chemical activity under stretching which is illustrated by a number of examples related tu double and triple carbon bonds. The chapter is organized as following. Section 2 opens the discussion by introducing basic main concepts in use. Main regularities concerning covalent bonds in light of stretching and breaking are considered in Section 3 addressing tetrel atoms from group 14 of the Mendeleev table. Section 4 is devoted to stretched covalent bonds in due course of chemical action. Mechanical stretching of covalent bonds is considered in Section 5. Conclusion summarises the discussed essentials.

## 2. Basic theoretical concept

When looking at quantities which are in operation within the framework of the UHF theory, effectively unpaired electrons seem to be the proper tool to describe stretching and breaking of chemical bonds quantitatively. The approach, first suggested by Takatsuka, Fueno, Yamaguchi (TFY) over three decades ago [17], was elaborated by Staroverov and Davidson later on [18]. As shown, the growth of internuclear distances between valence electrons, which provide the covalent bond formation, causes the appearing of effectively unpaired electrons since the electrons become correlated. The approach was firstly applied to the dissociation of $H_2$, $N_2$, and $O_2$ molecules [19] exhibiting the breaking of the relevant covalent bonds that accomplishes the bond stretching followed with bond's progressive radicalization.

The radical character of a molecule is commonly perceived as a one-electron property. Although an open-shell singlet has arguably more radical character than a closed-shell species, the difference is not evident from conventional one-electron distributions. Indeed, the total charge density $\rho(r)$ by itself contains no implication of unpaired electrons, whereas the exact spin density $\rho_u(r) = \rho_\alpha(r) - \rho_\beta(r)$ for a singlet is zero at every position. To exhibit unpaired electrons, TFY suggested new density function

$$D(r|r') = 2\rho(r|r') - \int \rho(r|r'')\rho(r''|r')dr'' \qquad (1)$$

that exhibits the tendency of the spin-up and spin-down electrons to occupy different places in space. The function $D(r|r')$ was termed the distribution of 'odd' electrons, and its trace

$$N_D = trD(r|r') \quad (2)$$

was interpreted as the total number of such electrons [17]. The authors suggested $N_D$ to quantitatively manifest the radical character of the species under investigation. Two decades later Staroverov and Davidson changed $D(r|r')$ by the 'distribution of *effectively unpaired electrons*' [18, 19] emphasizing a measure of the radical character that is determined by $N_D$ electrons taken out of the covalent bonding. Even in the TFY paper was mentioned [17] that function $D(r|r')$ can be subjected to a population analysis within the framework of the Mulliken partitioning scheme. In the case of single Slater determinant Eq. (2) takes the form [18]

$$N_D = trDS, \quad (3)$$

where

$$DS = 2PS - (PS)^2. \quad (4)$$

Here $D$ is the spin density matrix $D = P^\alpha - P^\beta$, $P = P^\alpha + P^\beta$ is a standard density matrix in the atomic orbital basis, and $S$ is the orbital overlap matrix ($\alpha$ and $\beta$ mark different spin directions). The population of effectively unpaired electrons on atom $A$ is obtained by partitioning the diagonal of the matrix DS as

$$D_A = \sum_{\mu \in A}(DS)_{\mu\mu}, \quad (5)$$

so that

$$N_D = \sum_A D_A. \quad (6)$$

Staroverov and Davidson showed [19] that the atomic population $D_A$ is close to the Mayer free valence index [20] $F_A$ in general case, while in the singlet state $D_A$ and $F_A$ are identical. Thus, a plot of $D_A$ over atoms gives a visual picture of the actual radical electrons distribution [21], which, in its turn, exhibits atoms with enhanced chemical reactivity.

In the framework of the UHF approach, the effectively unpaired electron population is definitely connected with the spin contamination of the UHF solution state caused by single-determinant wave functions which results in a straight relation between $N_D$ and square spin $\langle S^2 \rangle$ [21]

$$N_D = 2\left(\langle S^2 \rangle - \frac{(N^\alpha - N^\beta)^2}{4}\right), \quad (7)$$

where

$$\langle S^2 \rangle = \left(\frac{(N^\alpha - N^\beta)^2}{4}\right) + \frac{N^\alpha + N^\beta}{2} - \sum_i^{N^\alpha}\sum_j^{N^\beta}\left|\langle \phi_i|\phi_j\rangle\right|^2. \quad (8)$$

Here $\phi_i$ and $\phi_j$ are atomic orbitals; $N^\alpha$ and $N^\beta$ are the numbers of electrons with spin α and β, respectively.

If UHF computations are realized via the *NDDO* approximation (the basis for AM1/PM3 semi-emiempirical techniques) [22], a zero overlap of orbitals in Eq. (8) leads to $S = I$, where $I$ is the identity matrix. The spin density matrix $D$ assumes the form

$$D = (P^\alpha - P^\beta)^2. \qquad (9)$$

The elements of the density matrices $P_{ij}^{\alpha(\beta)}$ can be written in terms of eigenvectors of the UHF solution $C_{ik}$

$$P_{ij}^{\alpha(\beta)} = \sum_{k}^{N^{\alpha(\beta)}} C_{ik}^{\alpha(\beta)} * C_{jk}^{\alpha(\beta)}. \qquad (10)$$

Expression for $\langle \hat{S}^2 \rangle$ has the form [23]

$$\langle \hat{S}^2 \rangle = \left( \frac{(N^\alpha - N^\beta)^2}{4} \right) + \frac{N^\alpha + N^\beta}{2} - \sum_{i,j=1}^{NORBS} P_{ij}^\alpha P_{ij}^\beta. \qquad (11)$$

This explicit expression is the consequence of the Ψ-based character of the UHF approach. Since the corresponding coordinate wave functions are subordinated to the definite permutation symmetry, each value of spin $S$ corresponds to a definite expectation value of energy [24]. Oppositely, the electron density ρ is invariant to the permutation symmetry. The latter causes a serious spin-multiplicity problem for the UDFT schemes [25]. Additionally, the UDFT spin density $D(r|r')$ depends on spin-dependent exchange and correlation functionals only and cannot be expressed analytically [24]. Since the exchange-correlation composition deviates from one method to the other, the spin density is not fixed and deviates alongside with the composition.

Within the framework of the *NDDO* approach, the total $N_D$ and atomic $N_{DA} = D_A$ populations of effectively unpaired electrons take the form [26]

$$N_D = \sum_A N_{DA} = \sum_{i,j=1}^{NORBS} D_{ij} \qquad (12)$$

and

$$N_{DA} = \sum_{i \in A} \sum_{B=1}^{NAT} \sum_{j \in B} D_{ij}. \qquad (13)$$

Here $D_{ij}$ present matrix elements of the spin density matrix $D$.

$N_{DA}$ in the form of Eq. (13) actually discloses the chemical activity of atoms just visualizing the 'chemical portrait' of a molecule. It was naturally to rename $N_{DA}$ as *atomic chemical susceptibility* (ACS). Similarly referred to, $N_D$ was termed as *molecular chemical susceptibility* (MCS) [27]. Rigorously computed ACS ($N_{DA}$) is an obvious quantifier that highlights targets to be the most favorable for addition reactions of any type at each stage of the reaction thus forming grounds for *computational chemical synthesis* [28]. Firstly applied to fullerenes, the high potentiality of the approach was exemplified by fluorination [29] and hydrogenation [30] of fullerene $C_{60}$. An accumulating review is presented in Ref. [31]. Later on the approach was successfully applied to hydrogenation [32] and oxidation [33] of graphene.

Oppositely to UHF, UDFT does not suggest enough reliable expressions for either $N_D$ or

$N_{DA}$. The only known detailed discussion of the problem comparing UHF and UDFT results with CASSCF and multireference configuration interaction (MRCI) ones concerns the description of diradical character of the Cope rearrangement transition state [34]. When UDFT calculations gave $N_D = 0$, CASSCF, MRCI, and UHF calculations gave 1.05, 1.55, and 1.45 *e*, respectively. Therefore, experimentally recognized radical character of the transition state was well supported by the latter three techniques with quite a small deviation in numerical quantities while UDFT results just rejected the radical character of the state. Serious UDFT problems are known as well in the relevance to $\langle \hat{S} \rangle^2$ calculations [35, 36]. These obvious shortcomings of the UDFT approach might be a reason why UDFT calculations of this kind are rather scarce.

Analysis of the MCS behavior along the potential energy curve of diatomic molecules, firstly performed by Staroverov and Davidson [18] and then repeated by Sheka and Chernozatonskii in [37], led to the idea of using the $N_D(R)$ dependence for the quantitative description of chemical bonds upon dissociation. As shown, a characteristic *S*-like character of the dependence is common for all the molecules. Thus, each *S*-curve involves three regions, namely, (I) $R \leq R_{cov}$, (II) $R_{cov} \leq R \leq R_{rad}$, and (III) $R \geq R_{rad}$. At $R<R_{cov}$, $N_D(R)=0$ and $R_{cov}$ marks the extreme distance that corresponds to the completion of the covalent bonding of the molecule electrons and which exceeding indicates the onset of the molecules radicalization that accompanies the bond breaking. $R_{rad}$ matches a completion of homolytic bond breaking followed by the formation of two free radicals with practically constant value of $N_D(R) = N_D^{rad}$. The intermediate region II with a continuously growing $N_D$ value from zero to $N_D^{rad}$ exhibits a continuous build-up of the molecular fragments radicalization caused by electron extraction from the covalent bonding as the corresponding interatomic bond is gradually stretched. Thus, the $N_D(R)$ curve can be considered as a specific graph that quantitatively describes the state of the entire covalent bond dissociation path.

It seems quite reasonable that similar S-like curve should be expected for any chemical bond. The intention to test the hypothesis and to find out how rich information can be extracted from the $N_D(R)$ graphs has stimulated the performance of computational experiments, the results of which will be discussed below. Such a broad experiment was made possible due to use of the CLUSTER-Z1 codes of semi-empirical UHF calculations (see a detailed description of the codes in [38]). Following the reaction-coordinate format, two atoms of a pair indicating a particular covalent bond were taken out of the optimization procedure at each step of the elongation of the distance between them while the remaining atoms were optimized at each step. The elongation increment was of 0.05Å in general and of 0.01-0.02Å when some details were considered more scrupulously [39].

## 3. Covalent bonds in light of their stretching and breaking

### 3.1. $C \leftrightarrow C$ bonds

Bonds formed by two carbon atoms are the most rich in content and its general representation in the form $C \leftrightarrow C$ covers a set of traditionally matched single C-C, double C=C and triple C≡C bonds. The $N_D(R)$ graphs in Fig. 1 present a general view on the bond family on an example of the gradual dissociation of ethane, ethylene, and propyne molecules thus representing a continuous stretching and breaking of the corresponding bonds. As seen in the figure, all the studied $N_D(R)$ graphs are of S-like shape but significantly different. Thus, the single-bond graph is of one-step S-shape while for double and triple bonds S-like curves are evidently of two- and three-step, respectively. The number of steps evidently corresponds to the number of individual bonds involved in the relevant $C \leftrightarrow C$ bond. Each of the graphs starts by a horizontal line corresponding to $N_D=0$, which evidences the absence of effectively unpaired electron since all electrons are covalently bound. The left-hand edge of the region corresponds to the equilibrium length of the bond $R_{eq}$ while the right-hand edge indicates at which interatomic distance the covalent bonding is violated thus pointing to the largest covalent bond length $R_{cov}$ to be reached, on one hand, and, on the other hand, from which the covalent bond can be considered as broken. This region can be characterized by both absolute and relative width $W_{cov} = R_{cov} - R_{eq}$ and $\delta W_{cov} = W_{cov}/R_{eq}$. Superscripts *sg*, *db*, and *tr* in text below are used to distinguish different bonds of the $C \leftrightarrow C$ set.

When reaching $R_{cov}$, each of the three $C \leftrightarrow C$ graphs undergoes a jump that indicates the beginning of the bond radicalization when breaking. The radicalization gradually proceeds while the interatomic distance increases, although quite differently for the three bonds. Thus, the radicalization of the C-C bond of ethane, started at $R_{cov}^{sg}$ =2.11Å is fully completed at $R \leq$ 3Å and two single radicals are formed. The radicalization of the C=C bond of ethylene starts at $R_{cov}^{db}$ = 1.38Å and is saturated at the same region as for the single bond at $R \leq$ 3Å where a pair of two-fold radicals is formed. However, on the way to a completed radicalization a clearly seen kink on the $N_D(R)$ graph occurs. The kink critical point corresponds to $N_D \approx 2e$ and, exhibited by differentiating, is located at $R_{k1}^{db}$ = 2.12Å that is well consistent with $R_{cov}^{sg}$ of the single bond. Therefore, the bond radicalization occurs in two steps, first of which is completed for a pair of π electrons by reaching $N_D \approx 2e$ while the second should be attributed to the dissociation of σ bond until $N_D \approx 4e$ is reached. The $N_D(R)$ graph of the C≡C bond of propyne, preserving a general S-like pattern, shows a two-kink behavior. As seen in Fig. 1, the bond radicalization starts at $R_{cov}^{tr}$ = 1.24Å and the first kink is located in the region of $N_D \approx 2e$ at $R_{k1}^{tr}$ = 1.40Å that is consistent with $R_{cov}^{db}$ = 1.38Å of the C=C bond of ethylene. In the region of $N_D \approx 4e$, the second kink is observed, whose critical point at $R_{k2}^{tr}$ = 2.10Å is consistent with $R_{cov}^{sg}$ of the C-C bond of ethane. A pair of three-fold radicals at $R \leq$ 3Å completes the bond breaking. Therefore, a gradual stretching of the C≡C bond of propyne can be presented as a consequent completed breaking and radicalization of the two pairs of π electrons first and then terminated by the radicalization of σ electrons followed with the total bond breaking. Consequently, one-step dissociation of the C-C bond of ethane is substituted with two- and three-step dissociation of the C=C bond of ethylene and C≡C bond of propyne, respectively.

Data presented in Fig. 1 allow speaking about a new aspect of chemical bonds concerning their radicalization. It should be remained that the radicalization is just a 'chemical' manifestation of the correlation of bond-involved valence electrons. From this viewpoint, single, double, and triple bonds are drastically different. Thus, the single bond is radicalized in the vicinity of its breaking that is smoothed due to radicalization. The smoothing makes the exact determination of the interatomic distance at with the bond is broken uncertain. Nevertheless, the differentiation of the $N_D(R)$ graph in this region reliably highlights $R_{cov}^{sg}$ as a clear singularity thus allowing its attribution to the fixation of the bond breaking. In the case of double C-C bond, $R_{cov}^{db}$ determines the breaking of π bond while $R_{k1}^{db}$, which coincide with $R_{cov}^{sg}$, fixes the breaking of σ bond. Similarly to the discussed, $R_{cov}^{tr}$ on the $N_D(R)$ graph marks the breaking of the first π bond while $R_{k1}^{tr}$ and $R_{k2}^{tr}$ fix the breaking of the second π bond and the remained σ bond, respectively. According to the observed consistency of $R_{k1}^{db}$ of ethylene and $R_{k2}^{tr}$ of propyne with $R_{cov}^{sg}$ of ethane, the latter value can be attributed to the interatomic distance at which any of the bonds of the discussed $C \leftrightarrow C$ set can be considered as broken.

Fixation of the bond breaking allows introducing such characteristic quantities as the absolute and relative width of the radicalization region $W_{rad}$ and $\delta W_{rad}$, respectively, that in the case of double and triple bonds of ethylene and propyne are of the form

$$W_{rad}^{db} = R_{k1}^{db} - R_{cov}^{db}; \qquad \delta W_{rad}^{db} = W_{rad}^{db} / R_{eq}^{db} \quad \text{for ethylene;} \tag{14}$$

$$W_{rad}^{tr} = R_{k2}^{tr} - R_{cov}^{tr}; \qquad \delta W_{rad}^{tr} = W_{rad}^{tr} / R_{eq}^{tr} \quad \text{for propyne.} \tag{15}$$

The corresponding sets of the $R_{eq}, R_{cov}, R_{k1}, R_{k2}, W_{cov}, \delta W_{cov}, W_{rad}, \delta W_{rad}$ data are listed in Table 1. As seen from the table, while $W_{cov}$ decreases when going from single to triple bond, $W_{rad}$ inversely increases. The feature is the main reason for a drastic difference in the chemical activity of the bonds which will be discussed in Section 4.

The $N_D(R)$ graph in Fig. 1 for N≡N bond looks quite similar to that of C≡C one. A sharp growth at 1.28Å definitely marks $R_{cov}^{tr}$ at which breaking of the first π bond starts. Still continuing sharp growth does not allow to fix the first kink that exhibits the breaking of the second π bond while the second kink, corresponding to the breaking of the remaining σ bond, is clearly observed at $R_{k2}^{tr}$=1.78Å. As expected, this point should coincide with $R_{cov}^{sg}$ of single N-N bond. Actually, $R_{cov}^{sg}$ determined for the N-N bond of hydrazine constitutes 1.80Å.

The presented single-determinant (SD)-CI-theory picture of the N≡N bond dissociation is well consistent with that obtained by using the multireference (MR) orbital entanglement-based analysis [13]. According to the latter, two π bonds are torn first and breaking of the remained σ bond, which occurs above 1.6Å, completes the cycle. Both conclusions are in full agreement with those followed from the $N_D(R)$ graph discussed above thus presenting that both single-determinant and multireference CI analysis are quite comparable. However, since characteristic singular points of the former approach is related to jumps and kinks while those of the latter one correspond to saturated regimes, the accuracy of the UHF approach in determining both intermediate and final bond breaking is evidently higher.

From the above it follows that any chemical bond should be described by a set of characteristic, only one of which, namely, the equilibrium length of chemical bond $R_{eq}$ can be standardized. However, empirical data show that $R_{eq}$ is characterized with a significant dispersion indicating the dependence of the quantity on surrounding atoms. From this viewpoint, the data presented for the considered three molecules may change when going to other atomic composition. Actually, data presented in Fig. 2 for single C-C bonds show that the absolute values of $R_{eq}^{sg}, R_{cov}^{sg}, W_{cov}^{sg}, \delta W_{cov}^{sg}$ sets (see Table 1) are different while the qualitative character of the relevant $N_D(R)$ graphs is conserved. Particularly, it should be noted that in polyatomic molecules the radicalization and breaking of single C-C bonds become more abrupt thus significantly narrowing the smoothing of the region of their radicalization.

The data on double C=C bonds, related to different molecules and presented in Fig. 3, show a common character of the relevant $N_D(R)$ curves with some difference of the $R_{eq}^{db}, R_{cov}^{db}, R_{k1}^{db}, W_{cov}^{db}, \delta W_{cov}^{db}, W_{rad}^{db}, \delta W_{rad}^{db}$ set values (see Table 1). All the $N_D(R)$ graphs have a two-step S-like image with a kink located in the region of $N_D$~ 2e. The kink critical points are well consistent with $R_{cov}^{sg}$ of the relevant single C-C bonds. The $N_D(R)$ data on C≡C bonds in polyatomic molecules have been so far absent.

## 3.2. C-O, C-H, and F-F bonds

Oxides and hydrides are the most popular species of the carbon chemistry that is why C-O and C-H bonds deserve a particular attention. The relevant $N_D(R)$ graphs presented in Fig.4 are related to the dissociation of single C-O bonds in ethylene glycol (-C-O) as well as C-C and C-H bonds of ethane. The $R_{eq}^{sg}, R_{cov}^{sg}, W_{cov}^{sg}, \delta W_{cov}^{sg}$ set of the bond parameters is listed in Table 1. As seen in the figure and follows from the table, the bond behavior is similar to that of single C-C one. The bond stretching is one-step involving an extended region of the elongated bond that is followed by the bond breaking and radicalization before a complete rupture. The elongation stage $\delta W_{cov}$ constitutes ~40-50% for all the bonds, while the radicalization smoothing of $R_{cov}^{sg}$ is small enough. As in the case of the C-C bond, one should expect a slight difference in the characteristics of C-O and C-H bonds depending on the atomic surrounding. Consequently, if $R_{eq}^{sg}, R_{cov}^{sg}, W_{cov}^{sg}, \delta W_{cov}^{sg}$ values for a certain single bond are the main goal of a study, $N_D(R)$ graphs should be calculated in each case separately, while the general character of the bond behavior is well reproduced by any of the calculated graphs.

Addition of the $N_D(R)$ graph related to F-F bond in the figure was stimulated by a possible comparison of the bond analysis by using either SR or MR approach of the CI theory. According to the former, F-F bond behave quite similarly to other single bonds: $R_{eq}^{sg}$ is well consistent with the standard value and the bond is broken at 1.60Å. The only thing that distinguishes it from the bonds formed by other atoms of the first raw of the Mendeleev table, is a relatively narrow $W_{cov}^{sg}$ range. However, as we shall see later, it is not an exclusively rare case. As for comparison with the data obtained by using MR approach of the CI theory [13], contrary to expectation, the agreement between SD and MR approaches is not good. Consistent in determining $R_{eq}^{sg}$, the MR approach discloses two stretched bonds at 2.50Å and 2.53Å and points to the bond breaking at 3.50Å. All the three quantities seem quite strange and unreasonable from the chemical viewpoint, which might point to some artifact due to exclusive dependence of the MR results on the choice of basic orbitals.

## 3.3. $X \leftrightarrow X$ covalent bonds of heavier tetrels (X=Si, Ge, and Sn)

Carbon is the first member of the tetrel family of group 14 atoms of Mendeleev's table and the outstanding importance of $C \leftrightarrow C$ bonds for organic chemistry is very stimulating for looking for a similar behavior of $X \leftrightarrow X$ chemical bonds formed by heavier tetrels. The similarity-and/or-unlikeness of different members of the family has been the content of hot discussions over a century [42]. The current around-graphene science represents a new milestone of activity in this direction and is full of suggestion of new prototypes of graphene foremost of which are based on the equivalence electron atoms such as silicon, germanium, and tin. Hexagon patterned one-atom-thick planar silicene, germanene, and stannen are importunately discussed (see Refs. [43-46] and references therein).

All the heavier tetrels lie below carbon due to which their covalent radii make a series 0.76-0.73-0.69; 1.11; 1.20 and 1.39Å for carbon ($sp^3$-$sp^2$-$sp^1$), silicon, germanium, and tin, respectively [47]. Obviously, the 4p-bond is gradually weakened, the influence of which on the behavior of $X \leftrightarrow X$ bonds is of high interest. To form a reliable platform for the comparative analysis of all the tetrels, the data presented below are related to molecules of the common structure, namely, ditetralanes $X_2H_6$, ditetrelenes $X_2H_4$, and ditetrylynes $X_2H_2$ ($C_2H(CH_3)$ in the case of carbon). All the heavier-tetrel molecules have been studied by now both theoretically and experimentally, albeit as embedded $X_2H_4$ and $X_2H_2$ bodies in rare gas solid matrices at low temperature due to high chemical reactivity (see Refs. [48-50] and references therein).

Figure 5 presents $N_D(R)$ graphs related to sets of $X \leftrightarrow X$ bonds for silicon, germanium, and tin species. When the equilibrium configurations of ditetralanes and ditetrelenes are not subjected to isomerism and are similar to those of ethane and ethylene, the ditetrylyne configurations had to be chosen among a large number of isomers. The available set of disilyne isomers is shown in Fig. 6 (see a detailed description of the configurations in Ref. [48]). Similar sets are characteristic for both germanium and tin species. As shown [48], a classical acetylene-like configuration $HX \equiv XH$ is not preferential by energy that is why trans-bent structures were chosen for the $N_D(R)$ graphs computations. The AM1-UHF calculations have confirmed that such a configuration corresponds to the equilibrium one of disilyne and distannyne while for digermyne the equilibrium configuration is close to the linear one.

As seen in Fig.5, the $N_D(R)$ graphs of the heavier tetrels behave quite similar to those shown in Fig.1 for carbon whilst shifted to longer interatomic distances. To make the comparison of all the tetrels more vivid, the data were accumulated for the total family in Fig. 7 for ditetralanes, ditetrelenes, and ditetrylynes separately. As seen in Fig.7a, the $N_D(R)$ graphs of tetralanes are one-step with clearly seen points $R_{cov}^{sg}$ whose values are listed in Table 1. The $R_{eq}^{sg}$ well correlate with doubled covalent radii given above while the $\delta W_{cov}^{sg}$ is twice less in average with respect to carboneous species. Completed sets of $R_{eq}^{sg}, R_{cov}^{sg}, W_{cov}^{sg}, \delta W_{cov}^{sg}$ values are given in Table 1.

The $N_D(R)$ graphs of tetrelenes in Fig. 7b are quite different. If, on the first glance, all the graphs demonstrate two-step radicalization, the first step is a reality for ethylene and distannene while in the case of disilene and digermanene it is abcent. The equilibrium interatomic distance $R_{eq}^{db} \approx 2.3$Å in both cases greatly exceeds $R_{cov}^{db}$ at 1.8 and 2.1Å for Si- and Ge-species, respectively. Two latter values were obtained computationally in due course of stepwise $R_{eq}^{db}$ contraction that is presented on the graphs by continuous curve without markers. In contrast to covalently saturated ethylene and distannene, equilibrium disilene and digermene are ~two-fold radicals. When proceeding with the bond elongation, the $N_D(R)$ graphs reveal kink in both cases that are positioned at $R_{k1}^{db}$, well consistent with $R_{cov}^{sg}$ in all cases, as seen in Table 1. Thus, equilibrium disilene and digermene, both with broken π bonds, continue their dissociation until breaking the remained σ bond at $R_{k1}^{db} \approx R_{cov}^{sg}$. Dissociation of distannene occurs quite similarly to that of ethylene described earlier. The relevant completed $R_{eq}^{db}, R_{cov}^{db}, R_{k1}^{db}, W_{cov}^{db}, \delta W_{cov}^{db}, W_{rad}^{db}, \delta W_{rad}^{db}$ sets are listed in Table 1.

The three-step radicalized $X \leftrightarrow X$ bonds are well presented in Fig. 7c, with two kinks well seen for digermyne and distannyne, particularly. However, only for propyne all the three steps are real. The equilibrium state of heavier tetrels is positioned much over $R_{cov}^{tr}$ in the region close to $R_{k2}^{tr}$ for disilyne and digermyne while near $R_{k1}^{tr}$ for distannyne as follows from Table 1 where the completed $R_{eq}^{tr}, R_{cov}^{tr}, R_{k1}^{tr}, R_{k2}^{tr}, W_{cov}^{tr}, \delta W_{cov}^{tr}, W_{rad}^{tr}, \delta W_{rad}^{tr}$ sets are presented. Therefore, in contrast to covalently saturated propyne, the other equilibrium tetrynes present ~4-fold radicals in the case of Si- and Ge-tetrynes while ~2.5-fold radical of Sn-tetryne, which means that both π bonds are broken in the first case while only one in stannyne. $R_{k2}^{tr}$ positions of all the species are well consistent with $R_{cov}^{sg}$ that determines interatomic distances at which all the contributors to the studied $X \equiv X$ bonds are broken.

The data presented in Fig. 7 are in good relation with common regularities known for tetrels. First, a close similarity is characteristic for Si- and Ge- based species. As known, the two tetrels of the same atomic composition are well interchangeable and highly intersoluble, both as molecules and solids (see a detailed discussion of the topic in Refs. [42, 48-50] and references therein). Second, the two tetrels significantly differ from both C- and Sn-ones [50]. Third, $X_2H_4$ and $X_2H_2$ in the case of Si and Ge are highly chemically active and can be fixed empirically only as rarely distributed embedded species in solid rare gas matrices at low temperature (see the relevant references in Ref. [50]). Evidently, the radical character of the species can explain such a behavior. If to focus on the peculiarities of the bond radicalization characteristic, two more corollaries can be made. The first concerns the difference of the radicalization behavior of Si- and Ge- tetrenes and tetrynes in details on the background of the common similarity of their $N_D(R)$ graphs. The second is related to the pairwise similarity of C- and Si- tetrenes and tetrynes as well as those of Ge- and Sn-tetrels with respect to the radicalization rate. According to the finding, certain likeness should be expected for the relevant atomic compositions that do or may involve the relevant $X = X$ and $X \equiv X$ bonds. To check this prediction, let us look at single-hexagon and multi-hexagon structures of the tetrel family atoms.

Figure 8 presents equilibrium structures of $X_6H_6$ molecules while Table 2 contains their structural and radicalization parameters. The benzene-like pattern is characteristic for all the molecules, absolutely flat in case of $C_6H_6$ and $Si_6H_6$ while somewhat out of planarity for $Ge_6H_6$ and $Sn_6H_6$. In the latter case, the benzene-like configuration is energetically less favorable (by 25%) comparing with the boat-like configuration shown in Figs. 8e-f. All the benzene-like molecules are characterized by the only bond length while there are two bond lengths in the boat-like $Sn_6H_6$ molecule. The equilibrium bond length $R_{eq}^{db}$ of the first two molecules coincide with $R_{eq}^{db}$ of the $X_2H_4$ molecules that is why, $R_{eq}^{db}$ of $Ge_6H_6$ and $Sn_6H_6$ molecules is in the vicinity of $R_{cov}^{db}$. This circumstance explains why the molecules are not radicalized ($N_D$=0) similarly to $C_6H_6$ for which $R_{eq}^{db} \cong R_{cov}^{db}$. Therefore, only in $Si_6H_6$ molecule all bonds are radicalized due to $R_{eq}^{db} > R_{cov}^{db}$ as well as two longer bonds of the boat-like $Sn_6H_6$ molecule for the same reason. Addressing again to the structure of molecules, one should pay attention to a considerable reduction of $R_{eq}^{db}$ in $Ge_6H_6$ with respect to $Ge_2H_4$ (co. Tables 1 and 2) and, conversely, the increase in $R_{eq}^{db}$ of $Sn_6H_6$.

Following this brief analysis of the structural and radical character of the $X_6H_6$ molecules, one finds both similarity and difference of the species at the basic level. Obviously, similarity inspires hope to get silicene, germanene and stannene as prospective new-material playground of the around-graphene science. The similarity excuses a voluntary choice of the majority of computationists to take the flat honeycomb structure of graphene as the basic tetrene models. At the same time, the difference between the molecules casts doubt on the soundness of the choice of basic model. Let us see how these concerns are valid.

Figure 9 presents the results of the optimization of the preliminary equi-structural honeycomb compositions of $X_{66}$ tetrenes. The configuration corresponds to a rectangular (5x5) fragment that involves five hexagons along armchair and zigzag directions, respectively. The equilibrium structures are presented in top and side projections. As seen in the figure, carbon and silicon compositions preserve the honeycomb structure, perfectly planar in the former case and of slightly violated planarity in the latter. In both cases, the $X = X$ bond length values are quite dispersed and occupy interval, the limit values of which are given in Table 2. The relevant intervals are exhibited in Fig. 7b by thick horizontal bars. The presence of the bonds, lengths of which exceed $R_{eq}^{db}$, provides a considerable total radicalization of the fragments ($N_D$) in both cases.

In contrast to the above species, $X_{66}$ tetrenes of germanium and tin do not preserve the honeycomb structure in due course of the optimization. Their initial structures in Fig. 9c and e are just replica of the equilibrium structure of $Si_{66}$ in Fig.9b. The followed optimization drastically disturbs the structures leaving only small clusters of condensed hexagon rings and making them considerably non planar. The bond lengths cover much wider interval, abandoned with short bonds, for which $R_{eq}^{db} < R_{cov}^{db}$ (see Fig. 7b). The latter explains why the total radicalization of both fragments is less then in the case of $C_{66}$ and $Si_{66}$ (see Table 2).

According to the data presented in Fig. 9 and Table 2, the total radicalization and violation of the honeycomb structure are main two reasons that greatly complicate the existence honeycomb structures of higher tetrenes in practice. The former is mainly related to $C_{66}$ and $Si_{66}$ fragments while the latter concerns $Ge_{66}$ and $Sn_{66}$. The radicalization of graphene and the answer to the question why it does not prevent from existing graphene under ambient conditions are considered in Refs. [14, 31, 51] in details. Briefly summarizing, the graphene radicalization is mainly concentrated on the circumference and thus is usually well inhibited by the termination of edge atoms. As for silicene, the termination of edge atoms is not enough to inhibit its high radicalization since the latter remains still high on the atoms in basal plane [31, 52, 53] due to which free standing one-atom thick silicene sheet cannot exist under ambient conditions. Experimental evidence of 'silicene' is related to the hexagon-patterned monolayers of silicon atoms on either Ir(111) or Ag(111) surfaces (see review [44] and references therein) chemically bound with substrates. The interatomic distances within the layer are well consistent with $R_{eq}^{sg}$ of disilane (see Table 1) justifying $sp^3$ hybridization of valence electrons of silicon atoms. As for germanene and stannene, data from Table 2 tell that the fragment radicalization is much lower than for graphene and can not be considered as the main difficulty for the species existence. However, until now none of numerous attempts to get either germanene or stannene in practice has been successful. Inability of the tetrene atoms to form a lengthy honeycomb structure is apparently the major deterrent due to which the formation of wished free standing germanene and stannene is not achievable. Perhaps, this obstacle might be overcome by the choice of a suitable substrate on which surface the adsorbed tetrels can form hexagon-patterned structures.

**4. Stretched bonds in covalent compounds**

**4.1. Quantum-chemical aspect of bond stretching**

According to the general consideration presented in the previous Section, the behavior of any covalent bond under stretching in $W_{cov}$ and $W_{rad}$ areas determine its fate on the way to a complete dissociation. Evidently, this behavior is different in the two regions so that the bonds with the preference of their stretching in either $W_{cov}$ or $W_{rad}$ will behave quite differently. As seen in Table 1, $W_{cov}$ dominates for single bonds while $W_{rad}$ presents the main region for double and triple bonds. Data presented in Table 1 show that for single bonds, $\delta W_{cov}$ covers a large interval that constitutes 40-75% of the initial bond length thus pointing to a possibility of their considerable elongation, while leaving the bond atoms chemically inactive. In contrast, $W_{cov}$ regions for double and triple bonds are very short, so that the bond elongation, which keeps the bond chemically inactive, is relatively small and does not exceed 5%. The main transformation of the bonds under stretching occurs in the $W_{rad}$ region and concerns their radicalization that, in its turn, generates the chemical activity of previously inactive bond atoms and enhances the latter the more the longer the bond becomes.

In today's chemistry one can find a large number of examples to support this conclusion. However, before proceeding to the illustration, it is necessary to say a few words about possible causes of bond stretching. The bond stretching is present in the chemical life not only in due course of mechanochemical reactions. Particular conditions of chemical reactions themselves as well as peculiar properties of reactants may cause changing in the equilibrium values of chemical bond lengths thus making them seem stretched. Therefore one can speak about chemical and mechanical stretching and we will adhere this definition below when considering specific examples.

When the mechanical stretching is considered as quite evident and is largely discussed, the chemical stretching has not yet received enough attention. This explains the astonishment shown by the detection of 'abnormally' long bonds, as in the case of single C-C [8] and C-O [9, 10] bonds. In fact, the elongated $R_{eq}^{sg}$ of 1.647 Å, 1.659 Å, and 1.704 Å for C-C bond [8] as well as of 1.54 Å [9] and 1.622 Å [10] for C-O bond fall in the first quarter of the bond $W_{cov}^{sg}$ regions as follows from Table 1 and Fig. 4 and is far from the bond dissociation. A few other examples will be given below. The choice was quite subjective, however, the author would like to believe that it highlights the problem under consideration full enough.

**4.2. Chemically stretched covalent bonds**

*4.2.1. Single bonds*

The existence of a region of elongated bonds, characterized by $W_{cov}^{sg}$, is the manifestation of a freedom that is given to atoms to adapt to different environments formed by surrounding atoms and bonds as well as to provide molecule's photoexcitation and ionization while keeping its integrity. The chemical environment greatly influences the formation of new chemical bonds and the best way to highlight this effect is to trace a consequent polyderivatization of complex molecules. In contrast to practical chemistry, for which any particular polyderivative of, say, fullerene $C_{60}$, is not always completely successful hard work, quantum chemistry may deal with a large family of possible polyderivatives much more easily, when, particularly, additional support is provided by a specific algorithm of polyderivative models construction. Quantum molecular theory of fullerenes suggests such an algorithm, which allows tracing the stepwise polyderivatization of the molecules quite successfully [31]. In particular, fullerene $C_{60}$ showed itself as an excellent platform to reveal changes in its geometry in due course of various polyderivation reactions. Moreover, it afforded ground for separate observations of the changes occurred with double-bond carbon core and single-bond additions.

*Polyderivatives of fullerene $C_{60}$.*

A computational stepwise hydration and fluorination of fullerene $C_{60}$ is described in details in Refs. [29, 30]. A complete family of hydrides and fluorides from $C_{60}$ to $C_{60}H_{60}$ and $C_{60}F_{60}$ was considered. Figure 10 presents a summarized view on the key features connected with chemical bonding. Figure 10a shows the evolution of the C-H bond formation when the $C_{60}$ hydration proceeds from $C_{60}H_{18}$ to $C_{60}H_{36}$, $C_{60}H_{48}$, and $C_{60}H_{60}$. As seen in the figure, the fullerene hydrides can not be characterized by the only standard $R_{eq}^{sg}$. In contrast, the value is greatly varied and shows an average gradual increase as the hydrogenation proceeds. Moreover, $R_{eq}^{sg}$ at the very beginning of hydrogenation exceeds the tabulated standard value of 1.09Å and its deviation from the standard achieves 5% for $C_{60}H_{60}$. The next important conclusion concerns a clearly seen weakening of the chemical bonding in the course of the hydrogenation. The final result concerns an evident standardization of C-H bonds for the $C_{60}H_{60}$ species reflecting a high $I_h$ symmetry of the molecule. The weakening of chemical bonding is well supported empirically, particularly by changing the frequencies of C-H stretchings in vibrational spectra of $C_{60}H_{18}$, $C_{60}H_{36}$, $C_{60}H_{48}$, and $C_{60}H_{60}$ molecules discussed more that once in the fulleranes' book [55]. The above features are characteristic for the addend bonding in all the $C_{60}$ polyderivatives [31]. The most convincingly it is shown in Fig.10b for $C_{60}$ fluorides. However, a similar behavior is typical for $C_{60}$ cyanides and aziridines [56].

In the course of fullerene polyderivatization, the double-bond carbon skeleton of the pristine molecules is being filled with single C-C bonds. These bonds of 1.57-1.48Å and 1.64-1.50 Å in length dominate for $C_{60}H_{48}$ and $C_{60}F_{48}$ species, respectively, as seen in Figs. 10 c and d, and are transformed in average standartized C-C bonds of 1.52Å and 1.58Å for $C_{60}H_{60}$ and $C_{60}F_{60}$ molecules, respectively. Similar bond elongation accompanies consequent cyanation and aziridization, just showing when the covalent coupling of the carbon skeleton of fullerenes becomes weaker, as in the case of $C_{60}$ fluorides.

An impressive contraction of double bonds of the skeleton is seen in the figures as well. Obviously, an aspiration to compensate a large stress of the skeleton of $C_{60}H_{48}$ and $C_{60}F_{48}$ species caused by the appearance of a large number of elongated single bonds results in the contraction of the remaining double bonds up to $R_{eq}^{db} \approx 1.320$ Å. The value becomes less than $R_{cov}^{db} = 1.395$ Å which promotes a complete inactivation of the bonds thus terminating further chemical reactions. This explains why $C_{60}H_{48}$ is the last product after which the $C_{60}$ fluorination is stopped. In the case of hydrides, the double bond contraction as well as per-step coupling energy terminates the family with $C_{60}H_{36}$ [31].

*Polyderivatization of graphene*

Another important example concerns the main issue of the modern chemistry devoted to graphene. Chemical modification is one of the hot topics of the graphene science aimed at finding controllable regulators of graphene properties, both chemical and physical (see one of numerous reviews on the matter [57] and references therein). However, the chemical modification in each case is a complicated polyderivatization whose regularities are as complex as in the case of fullerenes. Figure 11 illustrates the said above on example of single C-H bonds that are formed in the course of the graphene hydrogenation. Those were obtained computationally when considering stepwise hydrogenation of a rectangular (5x5) graphene fragment (graphene membrane, see Fig. 9a) that contains five benzenoid rings along both armchair and zigzag edges [32]. As known [14, 51], graphene polyderivatization starts at circumference edge atoms thus completed by the fragment 'framing' by some or other addends. In the case of hydrogen, the framing is two-step: the first step concerns monohydrogen framing while the second one proceeds as two-hydrogen one [32]. As turned out, the first step is characterized by the strongest chemical bonding (see graph 1 in Fig. 11) while the second one considerably weakens the bonding (graph 2 related to the insert at the left) that also becomes quite irregular. What happens later on depends on the polyderivatization conditions. If the graphene membrane is fixed over perimeter and its basal plane is accessible to hydrogen atoms from both sides, the gradual per-step hydrogenation is completed with the formation of a regular chair-like structure previously named as graphane [58] (bottom insert at the right). Herewith, all C-H bonds are standardized at 1.121Å (graph 3) and 1.127Å (graph 4) length for carbon edge and basal plane atoms, respectively. If the membrane is still fixed over perimeter but its basal plane is accessible to hydrogen from one side only, a bent canopy-like structure is formed in due course of hydrogenation (top insert at the right) followed with the expected changing in the C-H bond structure. As seen in Fig.11, the framing of edge carbon atoms remains unchanged (graph 2) as if there was no hydrogenation within the basal plane. In its turn, the basal plane hydrogenation is followed by much weaker and more irregular chemical bonding (graph 5). The bond length growth constitutes ~3% in average. However, two bonds 77 and 78 cannot be depicted within the chosen scale since their length exceed $R_{cov}^{sg}$ for C-H bond from Table 1 and they are associated with the hydrogen molecule (see the top insert at right) formed by two desorbed atoms (details of such a behavior is described in [32]). Two other hydrides, different from the considered ones, are formed when the graphene membrane is free standing. The C-H-bond presentation of the both is quite peculiar and convincingly revealing the difference in the chemical bonding occurred in the cases as well. Thus, the picture painted by chemical bonds is a highly informative source of detailed knowledge about delicate processes that accompany derivatization of complex molecules.

These examples only on a tiny fraction lift the veil over the profound of chemical transformations that take place through single bonds. Evidently, the latter could not be possible if the bond length were standard and fixed. Besides, since $W_{cov}^{sg}$ restricts the freedom of such transformations, the letter could not be possible as well if $W_{cov}^{sg}$ were small. Chemists usually well understood this and intuitively accepted a considerable elongation of the bonds. The only question remained concerns the elongation limit. In practice, the majority of researchers rely upon the upper limit of the lengths embedded in widely used programs that are aimed at molecule imaging. If these limit values are much lower than $R_{cov}^{sg}$, a lot of chemically bonded compositions should be considered as consisting of separated parts. If the relevant $R_{cov}^{sg}$ values were inserted in the programs, a lot of atomic compositions with elongated chemical bonds would be found, so that the heralded 'abnormal' C-O [9, 10] and C-C [8] would cease to be a curious exception.

### 4.2.2. Double and triple bonds

Double and triple covalent bonds are prerogative of tetrel-based compounds. As follows from Table 1, the bond main specificity consists not only in small $W_{cov}^{db}$ and $W_{cov}^{tr}$ values but in their drastic decreasing when the number of atoms increases. Thus, say, for the benzene molecule $R_{eq}^{db}$ and $R_{cov}^{db}$ coincide so that stretching of any of the benzene C=C bonds occurs in the radicalization region and is followed with the appearance and further enhancement of the molecule chemical reactivity. It is this fact the reason why covalently saturated benzene is transformed into radicalized benzenoid units of fullerenes, carbon nanotubes, and graphene since their $R_{eq}^{db}$ values exceed $R_{cov}^{db}$ (this main peculiarity of the bodies is discussed in details in Refs. [14, 31, 51]).

As for simple molecules, alkenes and alkynes, which have in their structure one or more separated doubly and triply bonded pairs of atoms, dominate among other tetrenes and tetrynes. Fundamental organic chemistry tells us that alkenes are relatively stable compounds, but are more reactive than alkanes [59]. As for alkynes, their highly reactivity is a potential issue regarding their stability, use, and storage [60, 61]. The review [62] analyzes an emerging aspect in organic synthesis: the combination of alkynes and organocatalysis based on unique reactivity of alkynes. The data presented in Table 1 and Fig. 1 allow shedding light on the difference. As seen from the data, both alkene and alkyne is characterized by small $W_{cov}^{db}$ and $W_{cov}^{tr}$ values that are more than one order of magnitude less than $W_{cov}^{sg}$. The difference between alkene and alkynes is due to the scale of radicalization. As seen in Fig. 1, when a small stretching of double bonds causes radicalization up to ~1$e$, the same stretching of triple bonds evokes practically three-fold radicalization due to highly steep growth of the latter. This situation makes triple bonds extremely sensitive to stretching thus generating high radicalization thereby contributing to low stability of alkynes.

Chemical bonding, as was in the case of single bonds, evidently considerably influences the bond length, which in the case of alkenes and alkynes should be followed with enhance reactivity of the compounds formed. Above, the situation was described with respect to the transformation of inactive benzene molecule in considerably radicalized $sp^2$ nanocarbons such as fullerenes, carbon nanotubes, and graphene. Let us consider some examples related to alkyne-like bonding.

*Dymerization of p-diethylbenzene*

Figure 12 presents a view on what is happening when *p*-diethylbenzene (*p*-debz) is dimerized. The molecule presents a rare example when polymerization occurs in crystalline state caused by photoexcitation [63]. Evidently topochemical character of the solid photopolymerization is provided with parallel arrangement of the molecule benzene rings [64]. A detailed description of the further polymerization is discussed in [65]. Equilibrium structure of the *p*-debz monomer and dimer in Figs. 12 a and c are accompanied with the ACS maps of effectively unpaired electron distribution $N_{DA}$ (see Eq. (13)) over the molecules atoms shown in Fig. 12b and d, respectively. Addition of two acetylene units results in changing standard $R_{eq}^{db}$ of C=C bonds of the benzene molecule substituting the latter by two and four bonds of 1.392Å and 1.404Å in length, respectively. If $R_{eq}^{db}$ of the first two bonds even lie slightly below $R_{cov}^{db}$ =1.395Å characteristic for the pristine molecule, the $R_{eq}^{db}$ of four others exceeds the limit level thus promoting a remarkable radicalization of the molecule. Consequently, monomeric *p*-debz becomes 0.482-fold radical whose effectively unpaired electrons are distributed over the benzene ring atoms by ~0.07*e* at each (see Table 3), which is clearly seen in Fig. 12b. It should be noted that $R_{eq}^{tr}$ of both acetylene addends is kept below $R_{cov}^{tr}$ so that the units do not contribute into the molecule radicalization.

Dimerization causes a drastic reconstruction of the bond set thus promoting a large radicalization of the compound and lifting the monomer MCS $N_D$=0.482*e* to 8.474*e* for dimer. Chemical bonds of the benzene ring are still elongated forming three pairs of 1.394Å, 1.1.48Å, and 1.515Å in length, which results in lifting $N_{DA}$ as the indicator of their chemical reactivity to 0.27*e*, 0.28*e,* and 0.54*e*, respectively. However, the greatest changes concern the C$_1$ atoms of four acetylenes, for which previously zero $N_{DA}$ becomes equal to 0.94*e*. This is a result of the elongation of the C$_1$≡C$_2$ bonds to 1.334Å. As seen in Fig. 1, this new $R_{eq}^{tr}$ well explains the appearance of about one effectively unpaired electron per one bond. The revealed feature lets take a fresh look at remained a mystery over many years the local reactivity of polymers formed by molecules with acetylene groups [66]. Naturally, each individual case of such an activity deserves a separate consideration. But the overall trend is clear: polymerization strongly disturbs the relevant triple bonds causing their elongation thus promoting a drastic radicalization of the bond.

*Diphenylacetylenes, graphyne, and graphdyine*

Triple carbon bonds have recently become a hot spot in connection with a great desire to expand the number of materials related to graphene just meeting the increasing demand for carbon-based nanomaterials. Graphynes (GYs), consisting of benzenoid rings connected with chains of alkynes of different length, seem to be the most attractive (see a comprehensive review [67] that summarizes and discusses the state-of-the-art research of the issue, with a focus on the latest theoretical and experimental results). However, the main impression on promising interesting properties of GYs as well as on their possible applications has been provided by computational results while the experimental evidences are rather scarce. As for calculations, all of them were performed without taking into account a possible correlation of valence electrons of alkynes due to changing the relevant interatomic distances during GYs formation. Since the latter may be expected, let look at some basic components of GYs from this viewpoint.

A set of diphenylacetylenes (DPHAs) consisted of two phenyls connected with a varying number of acetylenes from 1 to 4 which are the simplest GY components is shown in the left panels of Fig. 13. The right panels of the figure display the relevant ACS maps that exhibit the presence of chemical reactivity of the molecules and present its distribution over the molecule atoms. Going from the top to bottom, one can see how the reactivity map changes in value and space when the acetylene group number increases. As seen in the figure, DPHA1 behaves quite similarly to *p*-debz discussed earlier. The inclusion of an acetylene between benzene rings causes a considerable elongation of some of the ring bonds thus promoting a significant radicalization of the rings as presented at the right-hand panel. The total number of effectively unpaired electrons $N_D$=1.22$e$ with fractional $N_{DA}$ values on the ring carbon atoms from 0.10$e$ to 0.08$e$ In contrast to the *p*-debz, a small radicalization of $N_{DA}$=0.06$e$ concerns the acetylene unit as well due to which the total $N_D$ value slightly exceeds the doubled value for *p*-debz . The summarized data are presented in Table 3. Inclusion one more acetylene unit between benzene rings promotes a remarkable elongation of both triple bonds, which, in its turn, results in the enhancement of their radicalization just lifting both $N_D$ and $N_{DA}$ values at the expense of the latter since characteristics of benzene rings remain practically unchanged. This trend is preserved with further increase of the number of triple bonds. As seen from Table 3 and Fig. 13, in process of growth of the bond number, the chemical reactivity of DPHAs is increasingly concentrated on the atoms of acetylene units evidencing the growing elongation of the latter. This might be explained by the transformation of a quite rigid single triple bond to a flexible chain of the bonds, which readily promotes the bond elongation.

The considered DPHAs lay the foundation of various GYs differing by the number of acetylene linkages between benzenoid rings. Thus, DPHA1 forms the ground of a carboneous material known as *graphyne* (*GY*), DPHA2 presents the basic element of *graphdyine* (*GDY*), and so forth [67]. Independently of a concrete structure of the involved DPHA, the composition like a six-petaled flower lays the foundation of the structure of ant GYs. Six-branched benzenoid ring determines each of the flower centers while another six rings terminate the flower petals. Each of these rings, one-branched previously, gradually becomes six-branched in due course of the GY growth in plane, which results in a particular triangle pattering of the GY body consisting of triangle closed cycles. Three benzenoid rings (rings below) form the vertices of the triangle while acetylene linkages (ligaments below) lie along its sides. Basing on these structural grounds, let us consider a consequent formation of a regular extended structure of *GDY*.

Figure 14 presents the formation of one- and two-triangle DPHA2-based compositions. The ACS maps, or chemical portraits, of the molecule at right-hand panels in the figure impressively exhibit changing in the atomic reactivity caused by changing in the composition structures that causes changing in bond lengths. As seen in the figure, the status of the ring ligament branching is the main reason of a sequential changing of both structure and reactivity of the compositions. The difference between one- and two-branched rings is evidently seen in Fig. 14a. It mainly concerns the rings themselves, two-branched of which is much highly reactive with respect to the one-branched one (see Table 3). The total MCS is distributed over rings and ligaments of such a way: 0.90$e$ is concentrated on two-branched ring while each one-branched ring takes 0.61$e$. Ligaments look much less reactive and quite identical and take 0.455$e$ each.

The addition of one more ligament to complete the triangle cycle leads to a considerable strengthening of the reactivity of the cycle as a whole, as seen in Fig. 14b. All the rings are now two-branched, which equally lifts their MCS to 0.95$e$. Thus, transition from one- to two-branched ring enhances its reactivity by ~55%. Apart from this, the ligament MCS increases as well up to 1.69$e$, equally to all of them thus enhancing their reactivity by ~25%. In this case, the asymmetry of the ACS distribution over the ring atoms caused by the addition of two ligaments to each ring is preserved.

A definite influence of a further branching on the ring reactivity is seen in Fig. 14c. Joining two triangles leads to a rhombic structure with two pairs of two- and three-branched rings, respectively. As seen in the figure and follows from Table 3, the MCS of both three-branched rings grows on 0.26$e$ which constitutes 27% of the reactivity of the preceding two-branched one, while the latter keep their MCS practically unchanged at 0.95$e$. Ligaments, which connect two-and three-branched rings, are characterized by the same MCS 0.60$e$ that only slightly exceeds 0.56$e$ corresponding to the connection between two-branched rings. However, the reactivity of the ligament connecting two three-branched rings increases up to 0.71$e$. The enhanced reactivity of three-branched rings as well as the connection between them is clearly seen in the ACS map presented in Fig. 14c.

Actually, six-branched benzenoid ring is the main motif of *GY*, *GDY* and any of their modifications. A completion of the ring branching is presented in Fig. 15. The data presented in the figure and in Table 4 well assist in tracing changes the concern both the main motif and ligaments on this way. Figure 15a presents pattern I related to four-branched central ring. As seen in the ACS map, the ring is highly selected from the surrounding due high MCS of 1.35$e$ on the background of much more modest one-branched rings and ligaments. The addition of one more ligament further enhances the effect lifting the central ring MCS up to 1.51$e$ (see Fig.15b) while only slightly influencing the remainder rings and ligaments. A completely branched ring is shown in Fig.15c. The ring MCS achieves it maximum at 1.64e at a completed branching that is accompanied with a slight increasing of the MCS of surrounding one-branched rings and ligaments only. Attention should be drawn on a high symmetry of the main motif pattern.

The above consideration of the successive branching of benzenoid rings of *GDY* recalls braiding Irish lace, whose main motive is presented in Fig. 15c. The analogy strengthens when looking at Fig. 16. where are given the patterns that gradually take us from individual motif to the regular arrangement of the latter on a broad cloth. The data presented in Table 4 will assist in getting a completed vision. The knitting is obviously a multi-stage complex process which is difficult to trace in all details. However, taking ACS maps as assistants it is possible to disclose general trend and regularities. Thus, two patterns exhibited in Fig. 16a and b alongside with the data in Table 4, which are schematically presented in Fig. 16 c, allow for making the following conclusions. 1. *GDY* presents a large cloth with a regular flower-like print where six-branched benzenoid rings play the role of the main floral motif while alkyne ligaments present thin twigs. 2. The motif is a radical but the status of its radicalization depends on surrounding structure. The highest ~7-fold (6.62-fold to be exact) radicalization is related to that one surrounded by six-branched benzenoid rings. 3. The motif main radicalization is concentrated on the ring while each ligament is about half less reactive. 4. The *GDY* cloth as a whole is highly radicalized and, consequently, chemically reactive. 5. The ACSs of the motif atoms are similar to those that are characteristic for carbon atoms of fullerenes, nanotubes and bazal-plane of graphene [31]. Similarly to the latter bodies, *GDY* can exist at ambient conditions, once inclined to a variety of chemical transformations. Cutting and saturation with defects will considerably enhance the body reactivity, which should be taken into account when discussing a possible controlling of electronic properties of *GDY* devices [67]. The conclusion might be important when planning practical *GDY*s applications.

## 5. Mechanical stretching of covalent bonds

### 5.1. Dynamic stretching

Discussed in the previous Sections has shown how deeply is the connection between the chemical reactivity and chemical bond skeleton of covalent compounds. This connection is of particular importance for double and triple bonding due to which an obvious conclusion follows: covalent bodies with double and triple bonds should show enhancement of their reactivity under mechanical loading. Empirical evidences of the effect have been observed rather occasionally and we shall refer to them in what follows. On the other hand, theoretical consideration was performed quite fundamentally, concerning, however, graphene only [68, 69]. Below we shall consider the main issues of the consideration on an example of the benzene molecule.

A detailed consideration of tensile deformation of the benzene molecule in the framework of the mechano-chemical reaction approach [70] is given in [68]. The approach consists in calculation of response in terms of energy, stress, as well as MCS and ACS ($N_D$ and $N_{DA}$) in due course of stepwise elongation of specific mechanical internal coordinates (MICs). Two such MICs, which join either 4-5 and 2-6 atom pairs or 5-6 and 1-2 atom pairs to provide either armchair (*ach*) or zigzag (*zg*) deformation of the molecule, respectively, are shown in Fig. 17a.

Figure 17 presents the elongation response of MCS and ACS related to the *ach* and *zg* deformation modes (Figs. 17c and d). As seen in the figure, from both structural and reactivity viewpoints the mechanical behavior of the molecule is highly anisotropic. Obviously, this feature is connected with the difference in the MIC atomic compositions related to the two modes, which results in the difference of the molecular fragments formed under rupture. In the case of *zg* mode, two MICs are aligned along $C_1$-$C_6$ and $C_3$-$C_4$ molecular bonds and two atomically identical three-atom fragments are formed under rupture. In the course of the *zg* mode, the MIC elongation is immediately transformed into the bond elongation. As shown in Section 3, for the C-C bond of the unstrained benzene molecule $R_{eq}^{db} \approx R_{cov}^{db}$ due to which the increment value of 0.05Å is significant enough for the unpaired electrons appearance even at the first step of elongation. According to Fig. 17b, the bonds' breaking occurs when the elongation achieves 0.75Å at both deformational modes, which corresponds to $R_{br}^{db}$ =2.15Å. The value is well consistent with $R_{k1}^{db}$ =2.14Å obtained for the benzene molecule earlier (see Table 1).

Both deformational modes consider the molecule breaking as a result of rupture of two C=C bonds. This explains high MCS values related to the final radicalization of broken pieces in both cases. However, the values are twice different for the two modes. This is explained by the difference of the breaking products. In the case of *zg* mode, the pristine molecule is broken into two identical $C_3H_3$ four-fold (3.5-fold to be exact) radicals while *ach* deformation results in the formation of $C_4H_4$ four-fold radical and inactive acetylene molecule.

Figures 17c and d present the reaction of different atoms on the molecule deformation. In the case of *ach* mode, the corresponding MICs connect atoms 1&5 and 2&6, respectively so that ~40% of the MIC elongation is transformed into that of two C-C bonds that rest on the MIC. This explains why $N_{DA}$ values on all carbon atoms are quite small in this case (Fig. 17c) until the MIC elongation $\Delta L$ is enough to provide the bond breaking. At the rupture moment, the acetylene molecule is slightly stretched, which explains the presence of unpaired electrons on atoms 5 and 6 (see Fig. 17c). However, a further relaxation of the molecule structure at larger elongation shortens the bond putting it below $R_{cov}^{tr}$ and unpaired electrons disappear. In the case of *zg* mode, both MICs coincide with C=C bonds that connect 2&4 and 1&6 atoms due to which the latter are deeply and equally involved in the deformation. Data presented in Fig. 17 convincingly show enhancement of the chemical reactivity of the object subjected to mechanical stretching.

These and other aspects of the deformation of benzene molecule impressively manifest the molecule mechanical anisotropy that lays the foundation of a drastic mechanical anisotropy of graphene [68]. Actually, the graphene deformation and rupture concern stretching and breaking of C=C bonds of its benzenoid units. Redistribution of the bonds in the graphene body at each step of deformation makes the latter extremely complicated and variable. Some general characteristics of the phenomenon are considered in [68, 69] in detalis. Just this situation has been recently implemented in practice [72] where a convincing evidence of the enhancement of chemical reactivity of graphene, subjected to tensile deformation, was obtained.

## 5.2. Static stretching

Besides dynamic, a number of static deformation modes exist. Within the framework of the issues raised in this chapter, of great interest are the effects of static stretching of systems with double and triple carbon bonds. In practice, such a situation has been realized for graphene related to the deformation of the carbon skeleton as roughnesses of different origin (wrinkles and bubbles). When the deformation causes stretching of the sheet skeleton, it is mandatory accompanied with enhancing the chemical reactivity. The effect was modeled by the chemically stimulated stretching of the graphene skeleton that can be traced by comparing those related to hydrides of the (5, 5) NGr molecule of the canopy-like and handbag-like ones obtained in the course of one-side hydrogenation of either fixed or free standing membrane, respectively [69]. As shown [14, 15], the skeleton bending causes increasing MCS from 31$e$ for a bare molecule to 46 $e$ and 54 $e$, for the canopy-like and basket-like skeletons, respectively.

The deformation-stimulated MCS rise leads to a number of peculiar experimental observations. Thus, if observed by HRTEM, the handbag-like skeleton might have look much brighter than the canopy-like one and especially than the least bright pristine molecule. In view of the finding, it is naturally to suggest that raised above the substrate and deformed areas of graphene in the form of bubbles, found in a variety of shapes on different substrates [72, 73], reveal peculiar electron-density properties just due to the stretching deformation. This explanation of a particular brightness of the bubbles at HRTEM images looks more natural than that proposed from the position of an artificial 'gigantic pseudo-magnetic field' [72].

The next observation concerns high-density wrinkles formed at a monolayer graphene structure grown on Pt(111) [74]. As shown, the wrinkles can act as nanosized gas-inlets in the graphene oxidation due to enhanced reactivity of wrinkles to oxygen. Analogous effect of enhanced reactivity was observed for monolayer graphene deposited on a Si wafer substrate, previously decorated with $SiO_2$ nanoparticles (NPs), and then exposed to aryl radicals [75]. As shown, the aryl radicals selectively react with the regions of graphene that covered the NPs thus revealing the enhanced chemical reactivity of the deformed graphene spots. The underneath substrate surface may be artificially configured with nanostructured grids of additionally deposited units, whereby the formed hilly graphene scarp may be used both *per se* and as a template for further regular chemical modification to tune the electronic properties in a wished manner.

## 6. Conclusion

Stretching and breaking of chemical bonds was considered in the current chapter from the viewpoint of correlation of valence electrons. In the framework of single-determinant CI theory effectively unpaired electrons lay the foundation of the quantitative measure of the correlation. The latter is related to *molecular chemical susceptibility* (MCS) determined by the total number of the unpaired electrons $N_D$ Single-determinant UHF computational schemes allow for determining this parameter quite reliably.

$N_D(R)$ plotting that describes the dependence of $N_D$ on the interatomic distance related to a selected chemical bond presents a graph of the bond behavior that starts at the equilibrium position of atoms $R_{eq}$ and is finished by a completed breaking that corresponds to the graph region of constant $N_D$ and is characterized by $\partial N(R)/\partial R \to 0$. However, the region is rather wide which makes the fixation of the bond breaking quite uncertain. At the same, the start of each breaking process is characterized by an abrupt changing at the graph due to which it can be fixed by the positions of the graph singularities characterized by maximum $\partial N(R)/\partial R$ values. In the case of single bonds, the singularity matches $R_{cov}^{sg}$, excess of which over $R_{eq}^{sg}$ determines the region $W_{\text{cov}}^{sg}$ of a possible stretching of the bond before breaking. $N_D(R)$ graphs of double bonds exhibits a successive breaking of π bond first, which is followed by the breaking of σ bonds. The transition between these two stages is characterized by a kink $R_{k1}^{db}$ that is attributed to the fixation of the double bond breaking. In the case of triple bonds, $N_D(R)$ graph shows two kinks caused by the π→π→σ sequences of breaking while $R_{k2}^{tr}$ is attributed to the breaking of the bond as a whole. $W_{\text{cov}}^{db}$ and $W_{\text{cov}}^{tr}$ regions of the latter two bonds are quite narrow which greatly limits the bond stretching that leave them inactive. On the contrast, the graphs show large regions $W_{rad}^{db}$ and $W_{rad}^{tr}$ for bond stretching before breaking that are characterized by enhanced chemical reactivity in the course of stretching.

The approach, applied to the analysis of the chemical bond behavior from different viewpoints, namely: 1) a comparative view on single, double, and triple carbon bonds; 2) the same but on these bonds in different surrounding; 3) the same but for heavier tetrel atoms, is well self consistent. A particular attention is given to the bond stretching. Stretching-caused peculiarities were considered for double and triple carbon bonds subjected to either chemical or mechanical action. Common in nature, the peculiarities are manifested differently for, say, p-diethylbenzene dimerization and a successive formation of graphdyine, on one hand, and uniaxial tension of the benzene molecule and, hence, graphene thus demonstrating a large scale of possibilities of the approach application.

Concluding, I would like to note that presented comprehensive analysis of chemical bond behavior was possible thanks to extended computational experiments performed by using semi-empirical version of the silgle-determinant UHF technique. I understand the pessimism of modern computationists in relation to semi-empirical methods of calculation. Is hard to access the methods almost of forty years antiquity on the background of such a dramatic development of quantum theory and the emergence of more and more new methods of calculation, which would seem to account for all the finest features of the behavior of electronic systems. However, first, we must not forget one of the Voltaire aphorisms that the best is the enemy of the good. Secondly, we should remember that any progress should to be paid. In our case, we are paying in the first place, a sharp reduction in the scale of computational experiment. Increasingly complex calculation methods require more calculation time and more powerful computing resources. Reducing the scale of the experiment, we miss the opportunity to see beyond individual calculations general patterns and trends that follow from them. Thus, of course, recognizing the great achievements made in the development of computational methods in the multireference CI theory, we should recognize the impossibility of obtaining currently using them the results presented in this chapter. At the same time, sophisticated semi-empirical methods, including the fundamentals of quantum chemistry of many-electron systems, enable a broad computational experimentation, bringing the results obtained from the sphere of individual tasks to the level of construction of general regularities. Examples for this are many, including, say, the latest one relating to the excited states of large molecules [76].

**Acknowledgements**
Author greatly appreciates financial support of the RSF grant 14-08-91376.

References

1. L. Pauling, *The Nature of the Chemical Bond*. Cornell University Press, 1960.
2. W. Locke (1997). Introduction to Molecular Orbital Theory. Retrieved May 18, 2005
3. N. H. March, *Electron Density Theory of Atoms and Molecules*. Academic Press, 1992.

4. R. F. W.Bader, *Atoms in Molecules - A Quantum Theory*. Oxford University Press, London, 1990.
5. J. L. Gazquez, J. M. del Campo, S. B. Trickey, R. J. Alvarez-Mendez and A. Vela, In *Concepts and Methods in Modern Theoretical Chemistry. Vol. 1. Electronic Structure and Reactivity*. Eds. S.K. Ghosh and P.K. Chattaraj. CRC Press, Taylor and Francis Group, Boca Raton, 2013, p. 295.
6. R.T.Sanderson, *Chemical Bonds and Bond Energy* Academic Press, New York, London, 1976.
7. B. deB. Darwent, *National Standard Reference Data Series*, National Bureau of Standards, No. 31, Washington, DC, 1970.
8. P.R. Schreiner, L. V. Chernish, P. A. Gunchenko, E. Yu. Tikhonchuk, H. Hausmann, M. Serafin, S. Schlecht, J. E. P. Dahl, R.M. K. Carlson and A.A. Fokin, *Nature*, 2011, **477**, 308.
9. M. Mascal, N. Hafezi, N. K. Meher and J. C. Fettinger, *J. Am. Chem. Soc*. 2008, **130**, 13532.
10. G. Gunbas, N. Hafezi, W. L. Sheppard, M. M. Olmstead, I. V. Stoyanova, F. S. Tham, M. P. Meyer and M. Mascal, *Nat. Chem*., 2012, **4**, 1018.
11. J. R. Lane, J.Contreras-García, J.-P. Piquemal and B. J. Miller, *J. Chem. Theory Comput*. 2013, **9**, 3263.
12. K.Boguslawski, P. Tecmer, Ő. Legeza and M.Reiher, *J. Phys. Chem. Lett*. 2012, **3**, 3129.
13. K. Boguslawski, P. Tecmer, G. Barcza, Ő. Legeza, M. Reiher, *J. Chem. Theory Comput*. 2013, **9**, 2959.
14. E.F.Sheka, *Int. J. Quant. Chem*. 2012, **112**, 3076.
15. E.F.Sheka., M. Hotokka et al. (eds.), *Advances in Quantum Methods and Applications in Chemistry, Physics, and Biology*, Progress in Theoretical Chemistry and Physics 27, Springer, Switzerland, 2013, 249.
16. I. Mayer, *J. Phys. Chem. A* 2014, **118**, 2543.
17. K.Takatsuka, T.Fueno and K.Yamaguchi. *Theor. Chim. Acta* 1978, **48**, 175.
18. V. N. Staroverov and E. R. Davidson, *Chem. Phys. Lett*. 2000, **330**, 161.
19. V.N. Staroverov and E.R. Davidson, *Int. J. Quant. Chem*. 2000, **77**, 316.
20. I. Mayer, *Int. J. Quant. Chem.* 1986, **29**, 73.
21. E.R. Davidson and A.E.Clark. *Phys. Chem. Chem. Phys*. 2007, **9**, 1881.
22. M.J.S.Dewar, and W.Thiel, *J. Am. Chem. Soc.* 1977, **99**, 4899.
23. D.A.Zhogolev and V.B.Volkov. *Methods, Algorithms and Programs for Quantum-Chemical Calculations of Molecules* (in Russian). Naukova Dumka, Kiev, 1976.
24. I.Kaplan, *Int. J. Quant. Chem*. 2007, **107**, 2595.
25. E. Davidson, *Int. J. Quant. Chem*. 1998, **69**, 214.
26. E.F.Sheka and V.A.Zayets, *Russ. J. Phys. Chem*. 2005, **79**, 2009.
27. E.F.Sheka, *J. Struct. Chem*. 2006, **47**, 593.
28. E.F.Sheka, *Int. J. Quant. Chem*. 2007, **107**, 2803.
29. E. F. Sheka, *J. Exp. Theor. Phys*. 2010, **111**, 395.
30. E. F. Sheka, *J. Mol. Mod*. 2011, **17**, 1973.
31. E.F.Sheka *Fullerenes: Nanochemistry, Nanomagnetism, Nanomedicine, Nanophotonics*. CRC Press, Taylor and Francis Group: Boca Raton, 2011.
32. E. F. Sheka and N.A.Popova, *J. Mol. Mod*. 2012, **18**, 3751.
33. E. F. Sheka and N.A.Popova, *Phys Chem Chem Phys*, 2013, **15**, 13304.
34. V.N. Staroverov and E.R. Davidson. *J. Am. Chem. Soc*. 2000, **122**, 186.
35. J.Wang, A.D. Becke, and V.H.Smith, Jr. *J. Chem. Phys*. 1995, **102**, 3477.
36. A.J.Cohen, D.J. Tozer and N.C.Handy, *J. Chem. Phys*. 2007, **126**, 214104 (4pp).
37. E.F.Sheka and L.A.Chernozatonskii, *J. Phys. Chem. C* 2007, 111, 10771.

38. P.K.Berzigiyarov, V.A.Zayets, I.Ya.Ginzburg, V.F.Razumov and E.F.Sheka, *Int.Journ.Quant.Chem.*, 2002, **88**, 441.
39. E. F. Sheka and N.A.Popova, *arXiv:*1111.1530v1[physics.chem-ph] 2011.
40. J.Kapp, M.Remko and P. v. R. Schleyer. *Inorg. Chem.* 1997, **36**, 4241.
41. N.W.Mitzel, U.Losehand, S.L.Hinchley and D.W.H.Rankin, *Inorg. Chem.*, 2001, **40**, 661.
42. S. Nagase, In: *The Transition State. A Theoretical Approach.* Ed. T. Fueno. Kodansha Ltd, Tokyo, Japan; Gordon and Breach Sci Pubs., Amsterdam, The Netherlands, 1999, 140.
43. S. Cahangirov, M. Topsakal, E. Akturk, H. Sahin and S. Ciraci, *Phys. Rev. Lett.*, 2009, **102**, 236804.
44. M. Xu, T. Liang, M. Shi and H. Chen, *Chem. Rev.* 2013, **113**, 3766.
45. G. Baskaran. *arXiv*: 1309.2242 [cnd-mat.str-el] 2013.
46. Y. Xu, B.Yan, H.-J. Zhang, J. Wang, G. Xu, P. Tang, W. Duan, and S.-C. Zhang, *Phys. Rev. Lett*. 2013, **111**, 136804.
47. B. Cordero, V. Gromez, A. E. Platero-Prats, M. Revres, J. Echeverrгıa, E. Cremades, F. Barragran and S. Alvarez, *Dalton Trans.*, 2008, 2832.
48. T. Shimizu. *Theoretical investigation of acetylene analogues of group 14 elements $E_2X_2$ (E=Si-Pb; X=F-I)*. archiv.ub.uni-marburg.de/diss/z2011/0043/pdf/dts.pdf, 2011.
49. P.Jutzi, U.Schubert (Eds.) *Silicon Chemistry. From the Atom to Extended Systems*. Wiley-VCH, Weinheim, 2003.
50. N. Wiberg. In *Silicon Chemistry. From the Atom to Extended Systems*. Wiley-VCH, Weinheim, 2003, p.85.
51. E.F.Sheka, *Int. J. Quantum Chem*. DOI: 10.1002/qua.24673, 2014.
52. E.F. Sheka, *arXiv*:0901.3663v1 [cond-mat.mtrl-sci], 2009.
53. E.F.Sheka, *Int. J. Quant. Chem*. 2013, **113**, 612.
54. R.Hoffmann, *Angew. Chem. Int. Ed*. 2013, **52**, 93.
55. F.Cataldo, S. Iglesias-Groth (Eds.) Fulleranes. The Hydrogenated Fullerenes. Springer, Berlin, Heidelberg, 2010.
56. E.F.Sheka, *J. Mol. Mod*. 2012, **18**, 1409.
57. U.N. Maiti, W. J. Lee, J. M. Lee, Y. Oh, J. Y.g Kim, J.E. Kim, J. Shim, T. H.Han and S. O. Kim, *Adv. Mater*, 2014, **26**, 40.
58. J.O. Sofo, A.S. Chaudhari and G.D. Barber, *Phys. Rev. B*, 2007, **75**, 153401.
59. L.G. Wade, (Sixth Ed.). *Organic Chemistry*. Prentice Hall, Upper Saddle River, NJ, 2006 p. 279.
60. P. Pässler, W. Hefner, K. Buckl, H. Meinass, A. Meiswinkel, H.-J. Wernicke, G.Ebersberg, R. Müller, J. Bässler, H.Behringer and D. Mayer, "*Acetylene*". *Ullmann's Encyclopedia of Industrial Chemistry*. Weinheim: Wiley-VCH. doi:10.1002/14356007.a01_097.pub3, 2008.
61. I. Beletskaya and C. Moberg, *Chem. Rev.* 1999, **99**, 3435.
62. R. Salvio, M.Moliterno and M.Bella, *Asian Journ. Org. Chem.* 2014, **3**, 340.
63. V.L.Broude, V.I.Gol'dansky, D.A.Gordon, *Zhurn. Fiz. Khimii*, 1968, **xx**, 345.
64. N.A.Ahmed, A.I.Kitaidorodsky and M.I.Sirota, *Acta Cryst.B* 1972, **28**, 2875.
65. E.F.Sheka and D.A.Gordon (to be published)
66. A.A.Berlin, G.A.Vinogradov, Yu.A.Berlin, Vysokomol. Sojed. A (in Russian) (Polym. Sci. A) 1980, 22, 862.
67. Y. Li, L. Xu, H.Liua and Y. Li, *Chem. Soc. Rev*. 2014, **43**, 2572.
68. E.F. Sheka, N.A.Popova, V.A. Popova, E.A. Nikitina, and L.Kh. Shaymardanova, *J. Mol. Mod.* 2011, **17**, 1121.
69. E.F.Sheka, V.A.Popova, N.A. Popova,.In: M. Hotokka et al. (Eds.), *Advances in Quantum Methods and Applications in Chemistry, Physics, and Biology, Progress in Theoretical Chemistry and Physics 27*, Springer, Switzerland, 2013, p.285.

70. E.A.Nikitina, V.D.Khavryutchenko, E.F.Sheka, H.Barthel and J.Weis, *J. Phys. Chem. A* 1999, **103**, 11355.
71. M.A. Bissett, S. Konabe, S. Okada, M. Tsuji and H. Ago, *ACS Nano* 2013, **7**, 10335.
72. N.Levy, S.A. Burke, K.L. Meaker, M. Panlasigui, A. Zettl, F. Guinea, A.H.d. Castro Neto and M.F. Crommie, *Science* 2010, **329**, 544.
73. N.Georgiou, L. Britnell, P. Blake, R.V. Gorbachev, A. Gholinia, A.K. Geim, C. Casiraghi and K.S. Novoselov1, *Appl. Phys. Lett.* 2011, **99**, 093103.
74. Y. Zhang, Q. Fu, Y. Cui, R. Mu, L. Jin and X. Bao, *Phys. Chem. Chem. Phys*., 2013, **15**, 19042.
75. Q.Wu, Y. Wu, Y. Hao, J. Geng, M. Charlton, S. Chen, Y. Ren, H. Ji, H. Li, D.W. Boukhvalov, R.D. Piner, C.W. Bielawski and R.S. Ruoff, *Chem. Commun*. 2013, **49**, 677.
76. K. Aryanpour, A. Roberts, A. Sandhu, R. Rathore, A. Shukla and S. Mazumdar, *J. Phys. Chem. C* 2014, **118**, 3331.

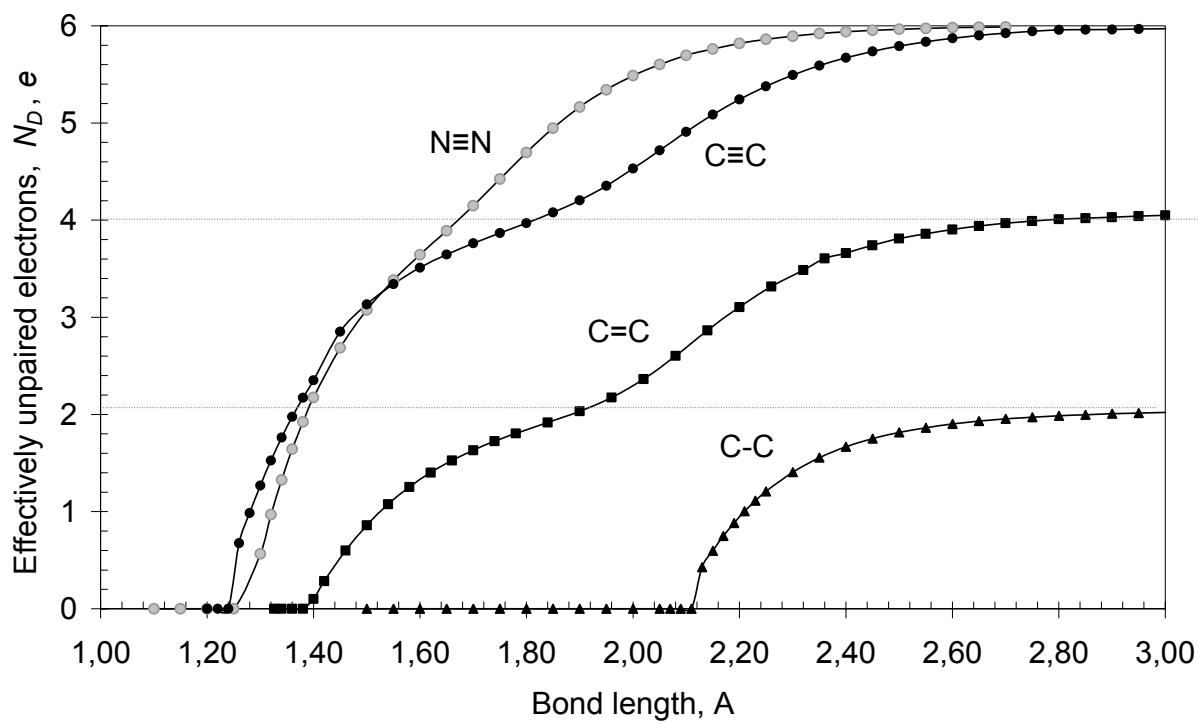

**Figure 1**. $N_D(R)$ graphs related to the dissociation of C-C (ethane), C=C (ethylene), C≡C (propyne), and N≡N (dinitrogen) bonds. AM1-UHF calculations.

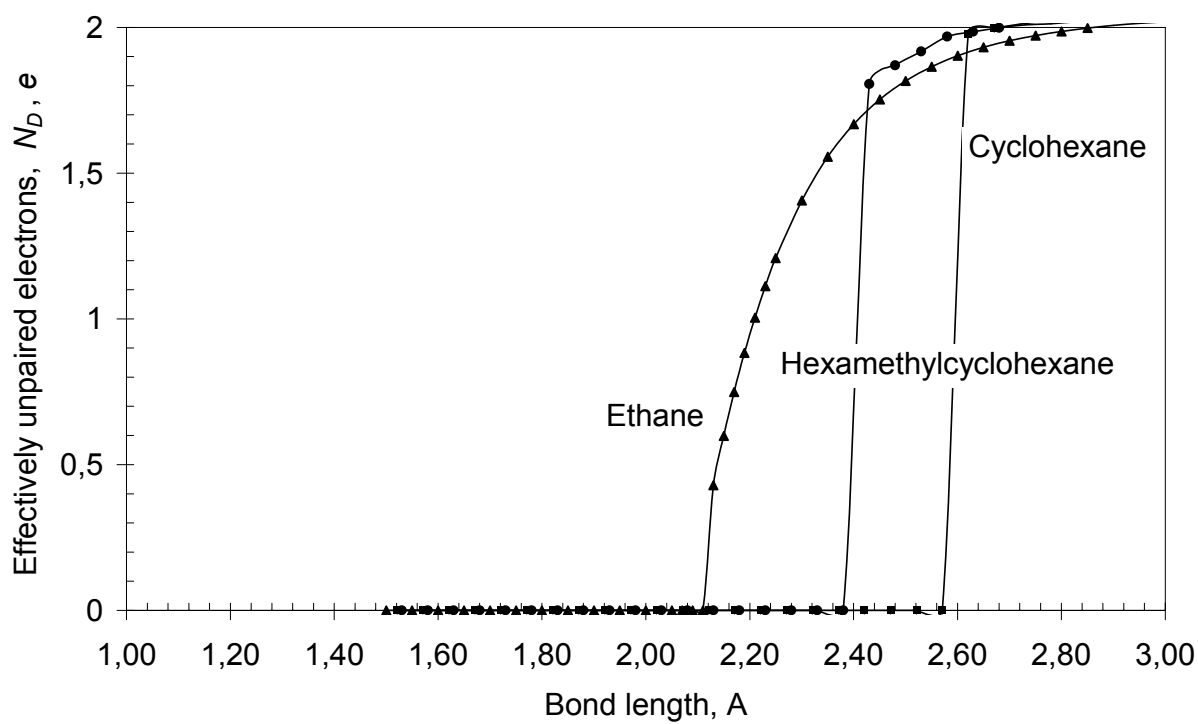

**Figure 2**. $N_D(R)$ graphs related to the dissociation of single C-C bonds in ethane, hexamethylcyclohexane, and cyclohexane. AM1-UHF calculations.

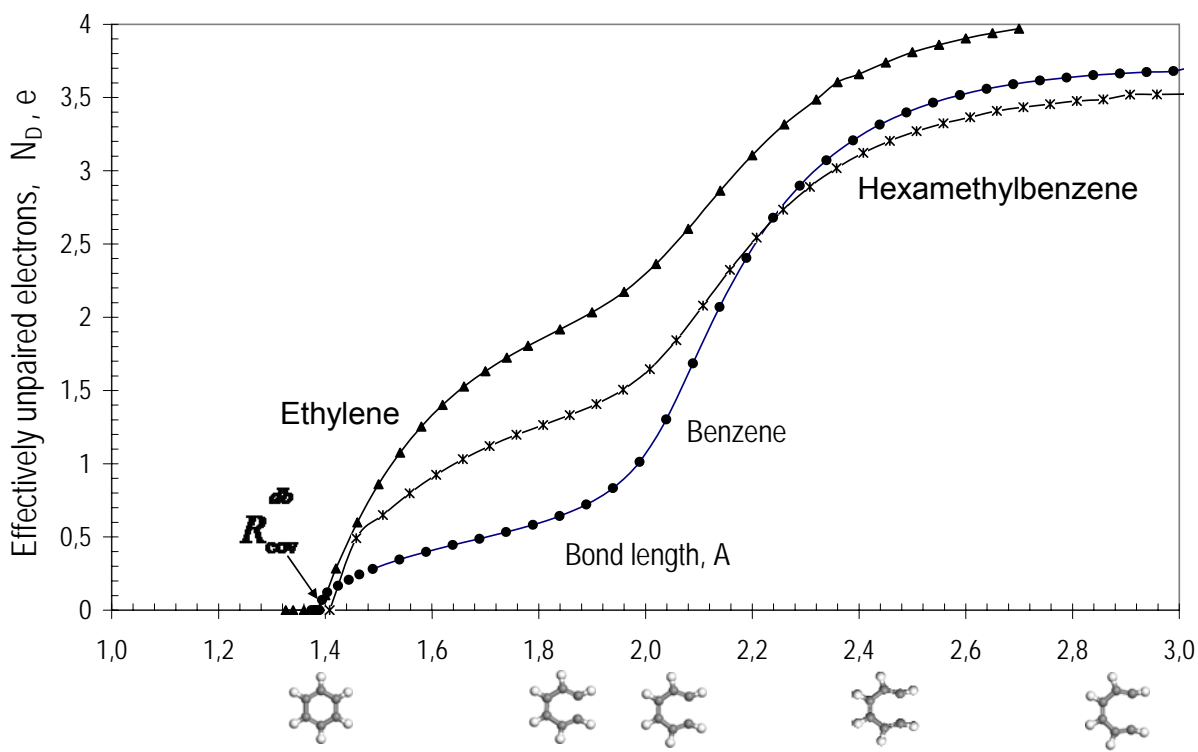

**Figure 3**. $N_D(R)$ graphs related to the dissociation of double C=C bonds in ethylene, benzene, and hexamethylbenzene. The inserted benzene structures are positioned correspondingly to the length of the right vertical C-C bond. AM1-UHF calculations.

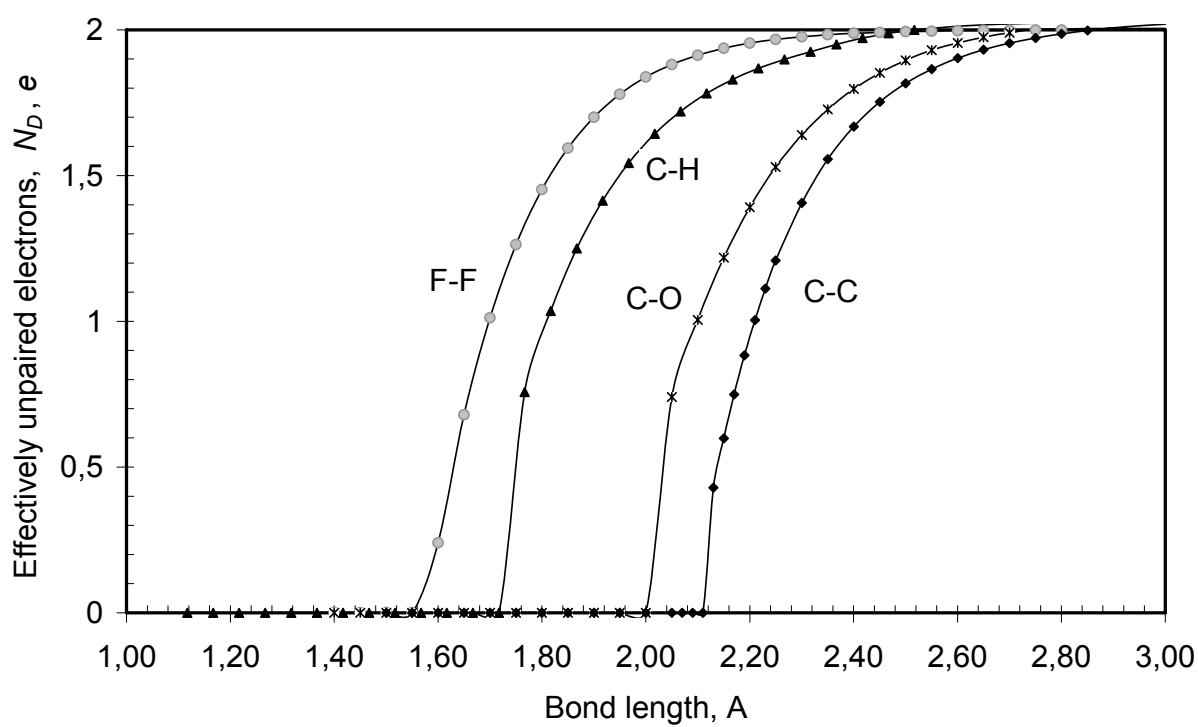

**Figure 4**. $N_D(R)$ graphs related to the dissociation of C-C and C-H bonds of ethane as well as C-O bond of ethylene glicol and F-F bond of fluorine molecule. AM1-UHF calculations.

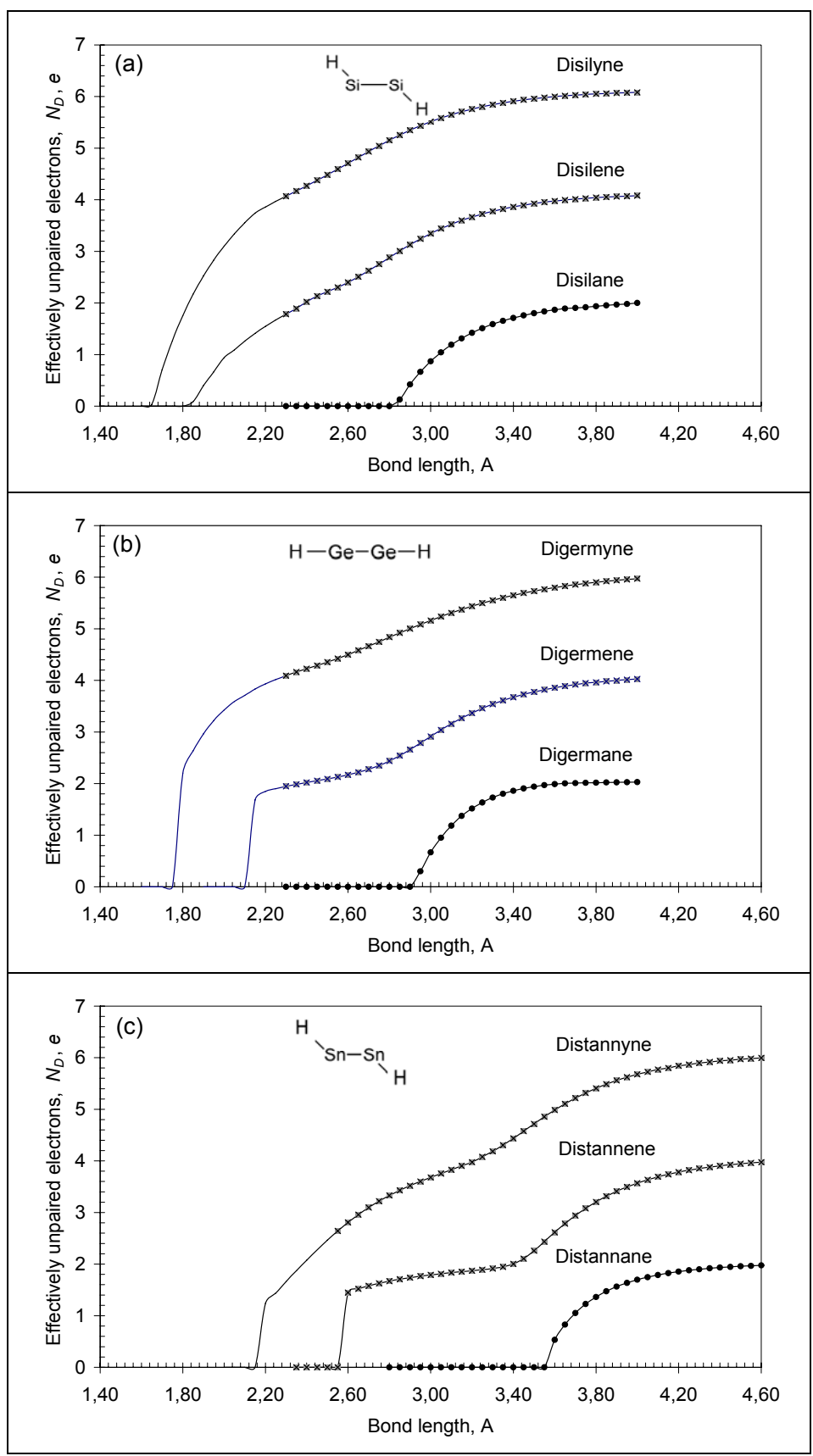

**Figure 5**. $N_D(R)$ graphs related to the dissociation of X-X, X=X, and X≡X bonds of X=Si (a) X=Ge (b) and X=Sn (c) tetrels. Plottings without markers correspond to the bonds' contraction below the relevant $R_{eq}$. Equilibrium structures of the molecules with X≡X bond under study are given as inserts. AM1-UHF for X=Si, Ge and PM3-UHF for X=Sn calculations.

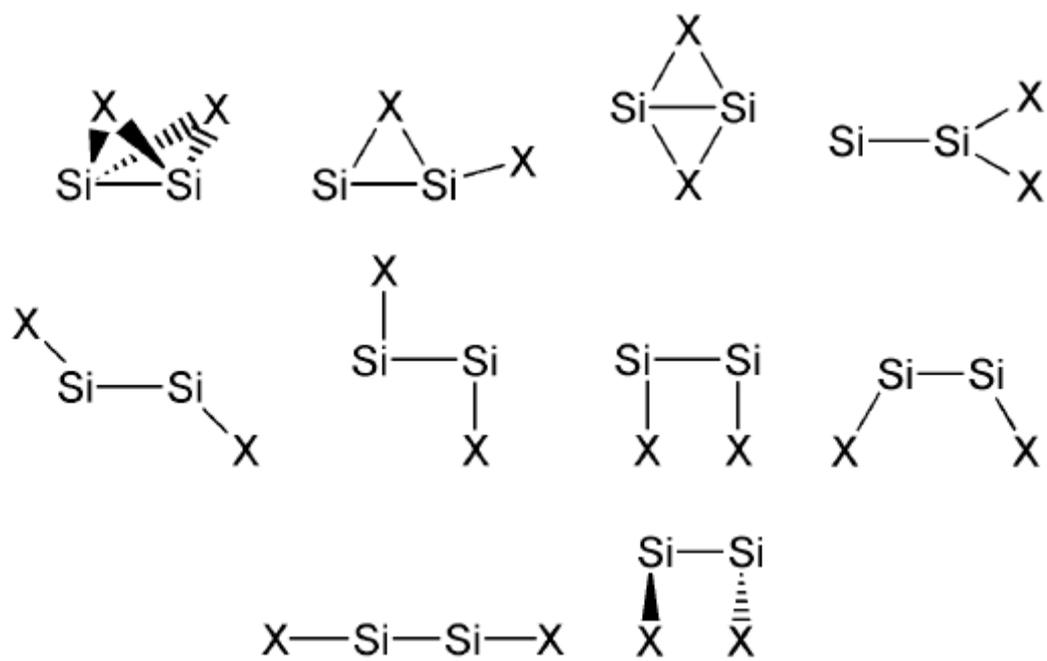

**Figure 6**. A set of possible isomers of ditrylynes (according to [48]).

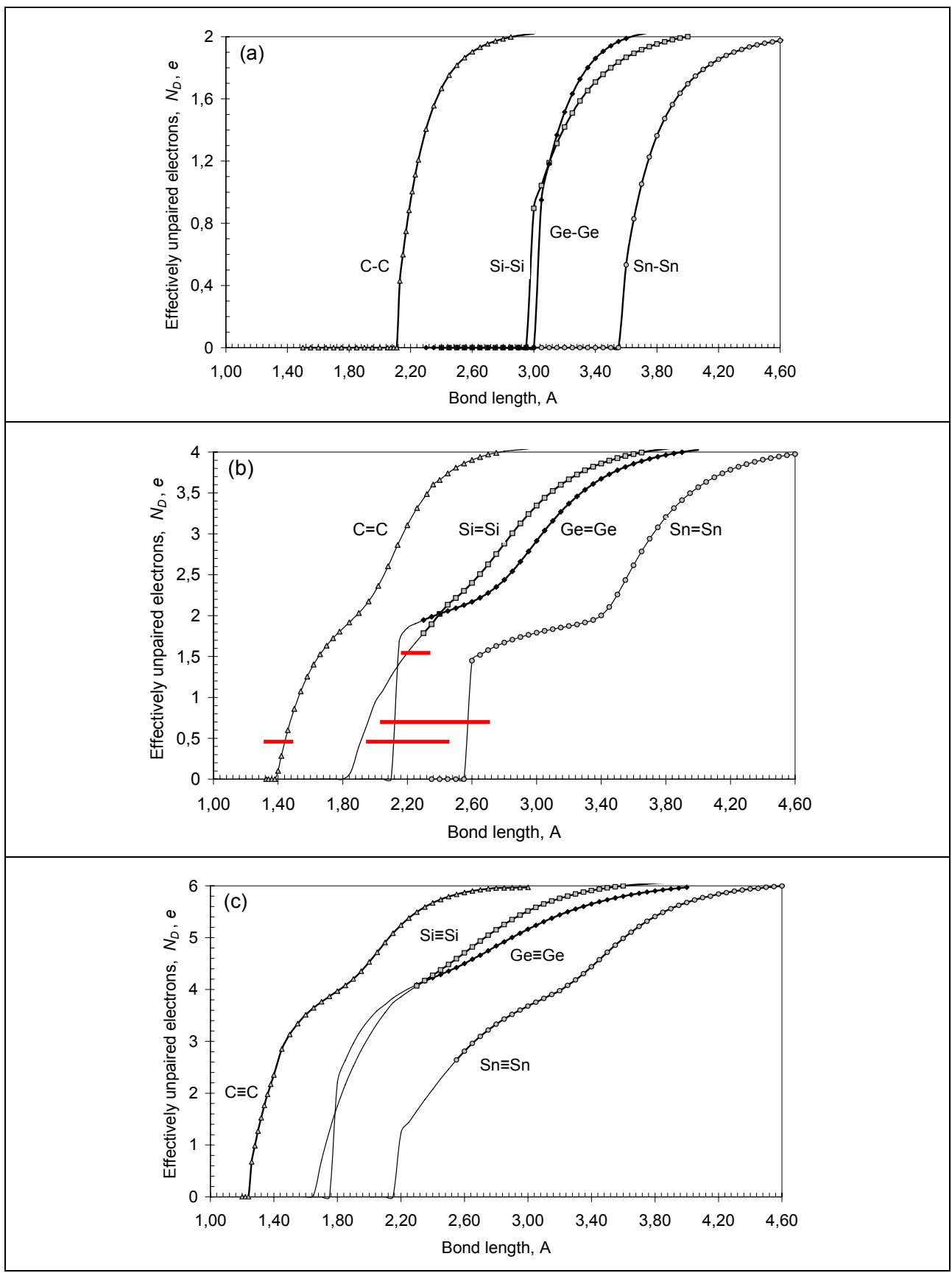

**Figure 7**. $N_D(R)$ graphs related to the dissociation of (a) X-X bonds; (b) X=X bonds (horizontal bars present the dispersion of the bond lengths of the relevant $X_{66}$ fragments (see Fig.9)); (c) X≡X bonds of the tetrels family; X=C, Si, Ge, and Sn. AM1-UHF for X=Si, Ge and PM3-UHF for X=Sn calculations.

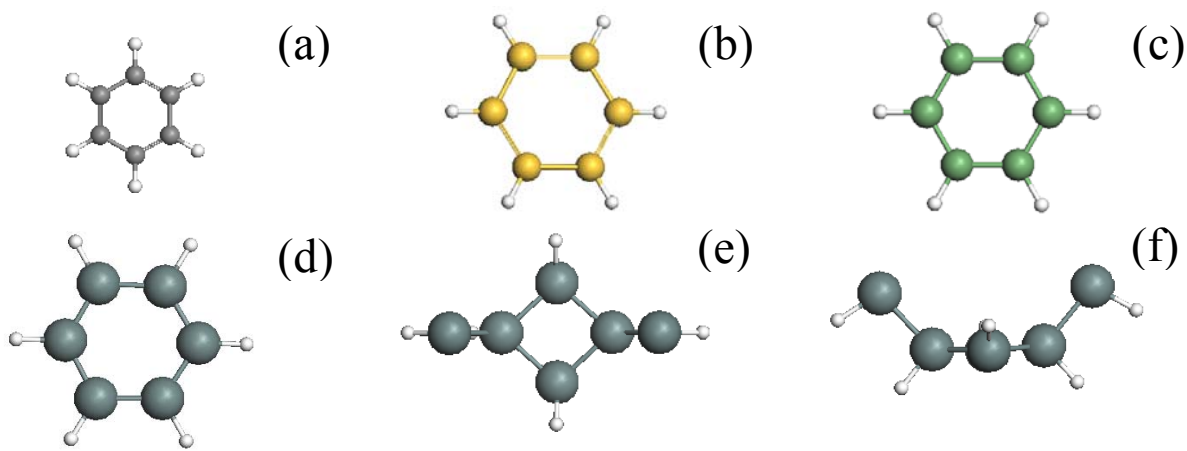

**Figure 8**. Equilibrium structure of $X_6H_6$ molecules when X=C (a), Si (b), Ge (c), and Sn (d-f). In the latter case, benzene-like (d) and boat-like (top (e) and side (f) views) compositions are presented. Gray, yellow, green, steel-gray, and white balls mark carbon, silicon, germanium, tin, and hydrogen atoms. The balls' sizes roughly correspond to the relevant van der Waals diameters. AM1-UHF for X=Si, Ge and PM3-UHF for X=Sn calculations.

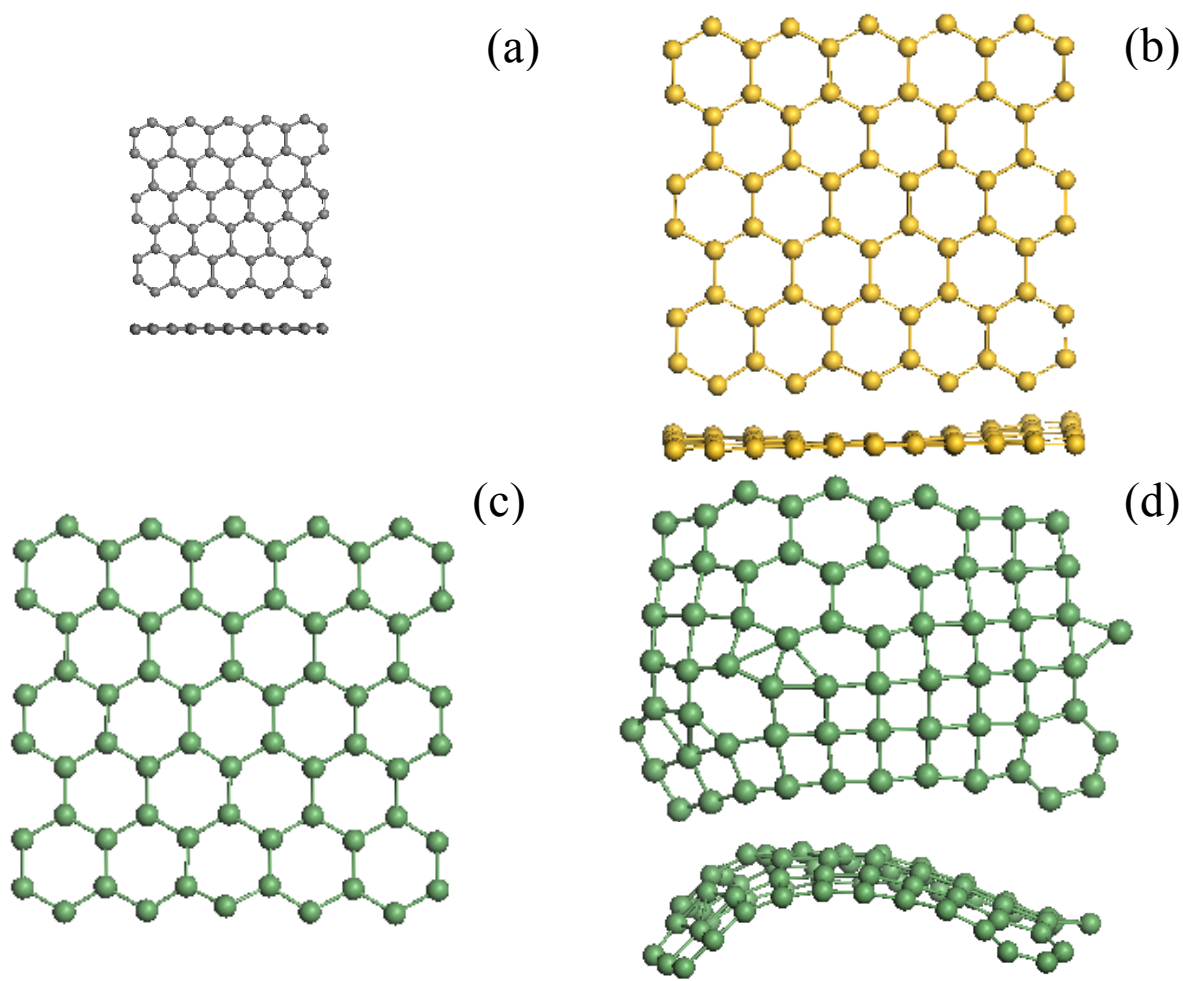

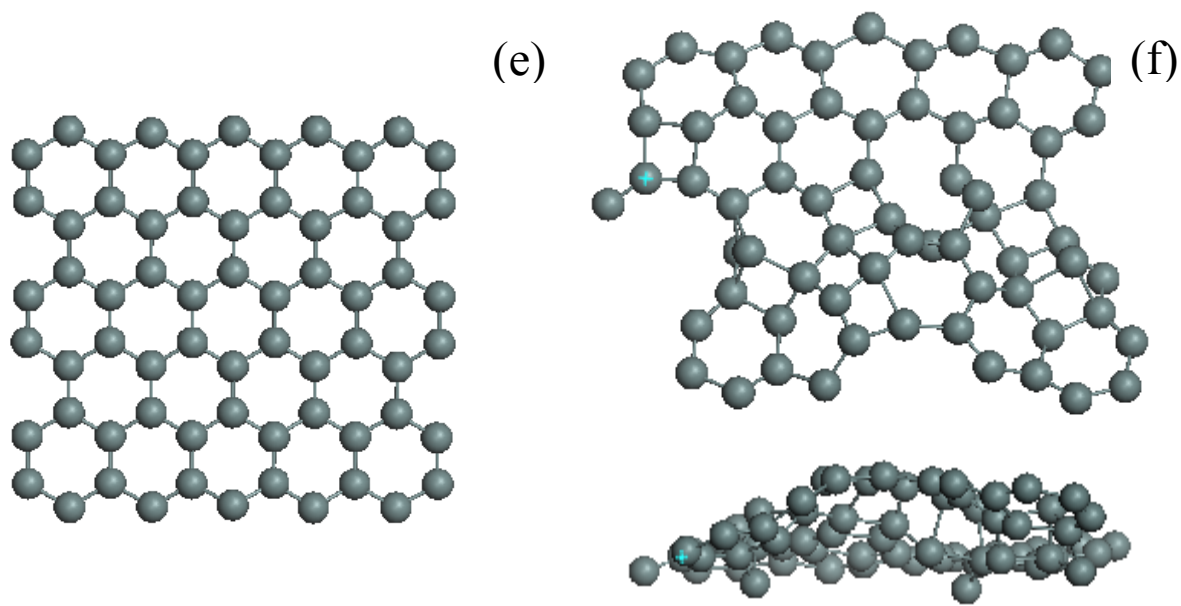

**Figure 9**. Equilibrium structure of $X_{66}$ (5x5) honeycomb fragments. (a) $C_{66}$; (b) top and side views of $Si_{66}$; (d) top and side views of $Ge_{66}$; (f) top and side views of $Sn_{66}$. (c) and (e) start $Ge_{66}$ and $Sn_{66}$ configurations, respectively. Atom marking see in the caption to Fig.8. The balls' sizes roughly correspond to the relevant van der Waals diameters. AM1-UHF and PM3-UHF calculations for X=C, Si, Ge and X=Sn, respectively.

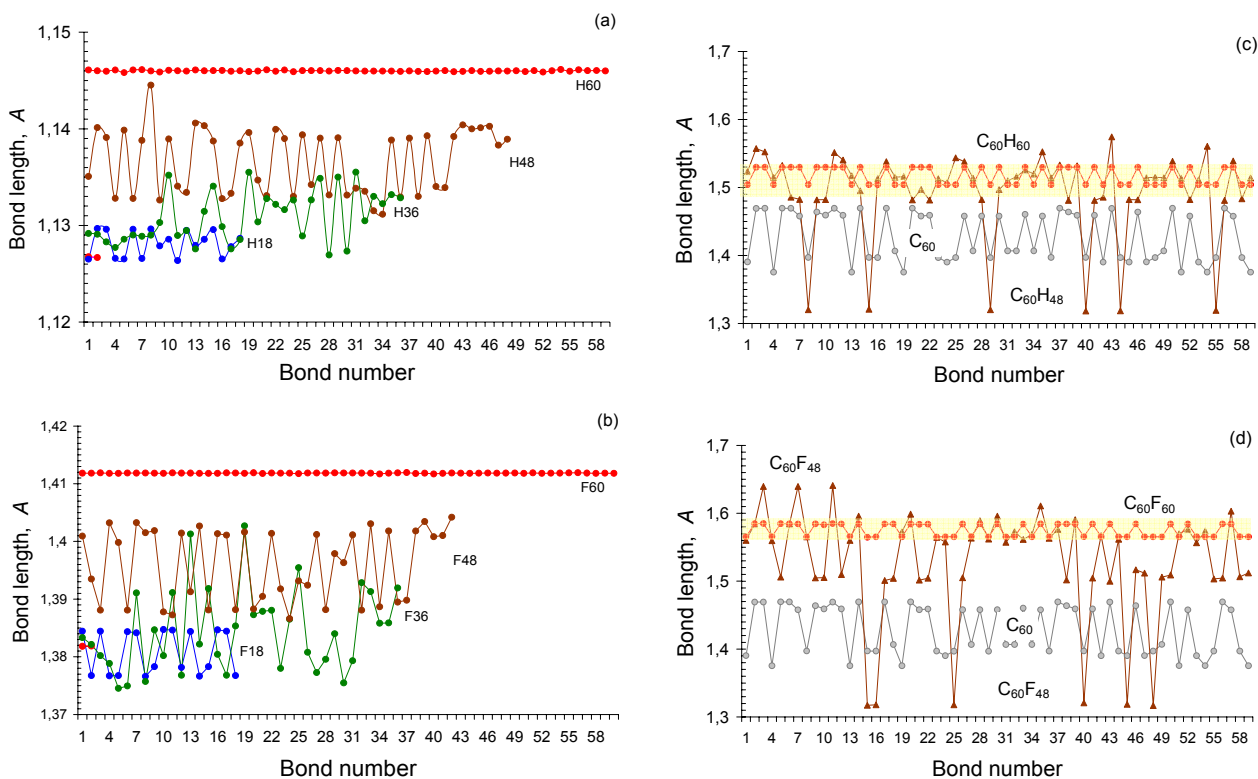

**Figure 10**. Covalent bonds of fullerene $C_{60}$ hydrides and fluorides. (a). C-H bonds of $C_{60}H_{18}$ (blue), $C_{60}H_{36}$ (green), $C_{60}H_{48}$ (brown), and $C_{60}H_{60}$ (red); (b) C-F bonds of $C_{60}F_{18}$ (blue), $C_{60}F_{36}$ (green), $C_{60}F_{48}$ (brown), and $C_{60}F_{60}$ (red); (c) C-C bonds of $C_{60}$ (gray), $C_{60}H_{48}$ (brown), and $C_{60}H_{60}$ (red); (d) C-C bonds of $C_{60}$ (gray), $C_{60}F_{48}$ (brown), and $C_{60}F_{60}$ (red). AM1-UHF calculations

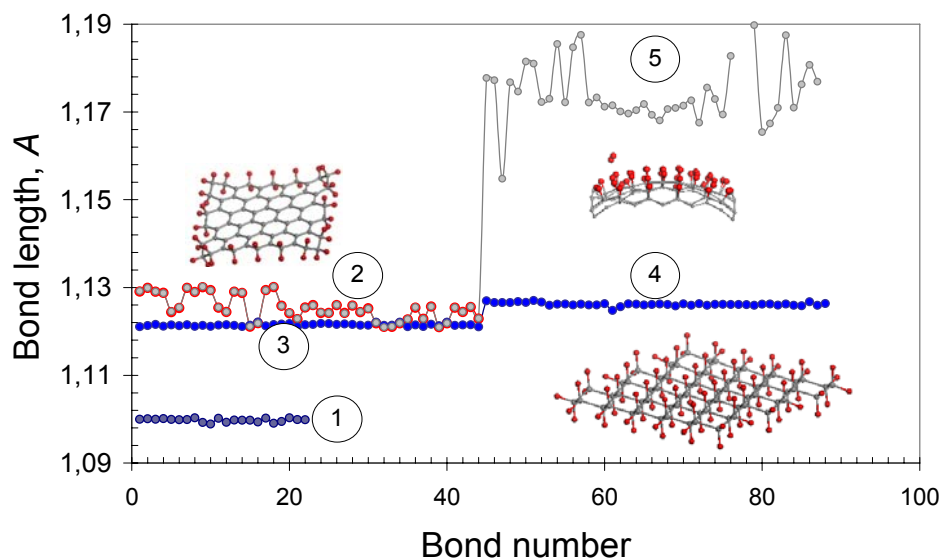

**Figure 11.** Covalent C-H bonds of graphene polyhydrides. (1). Monohydrogen framing of the (5x5) $C_{66}$ membrane; (2) Dihydrogen framing of the (5x5) $C_{66}$ fragment membrane by the left insert; (3) ibid as in (2) but related to regular chair-like graphane presented by the right bottom insert; (4) Monohydrogen covering of the graphane basal plane; (5) Monohydrogen covering of basal plane of fixed $C_{66}$ membrane accessible for hydrogen atoms from one side (the top insert r right). See detailed description in text. AM1-UHF calculations.

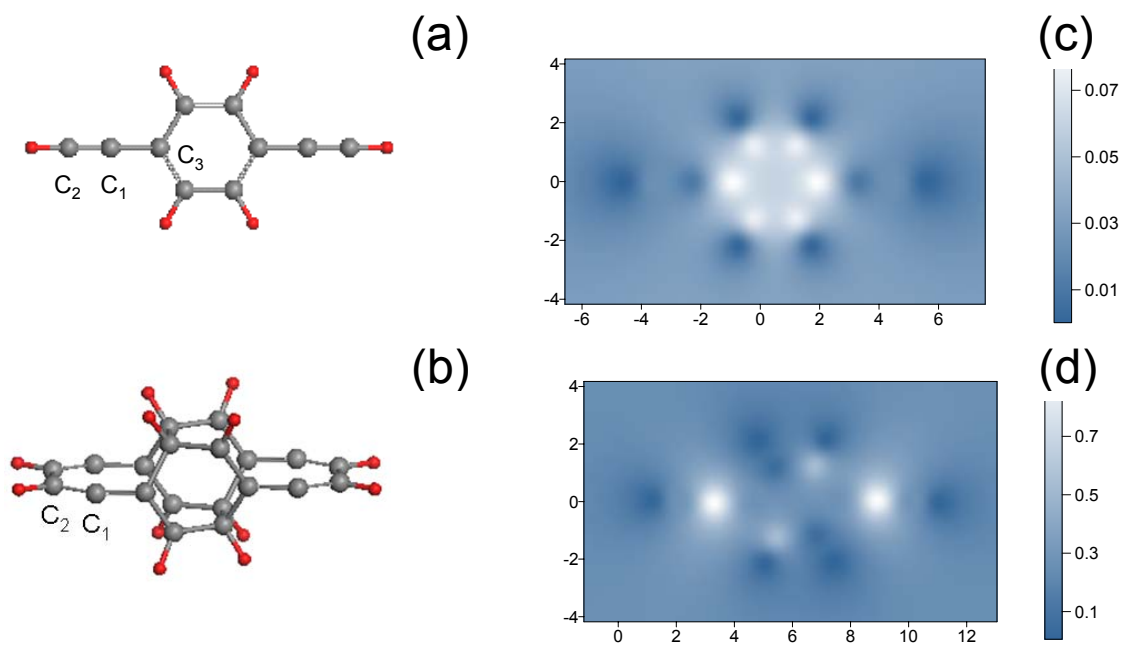

**Figure 12.** Equilibrium structures (left) and ACS ($N_{DA}$) maps (right) of monomer (a, c) and dimer (b, d) of *para*-diethylbenzene. $C_1$ and $C_2$ mark atoms of one of the acetylene units. Gray and red balls mark carbon and hydrogen atoms, respectively. The intensity scales in (b) and (d) differ by ten times. The maps axes are in Å. AM1-UHF calculations.

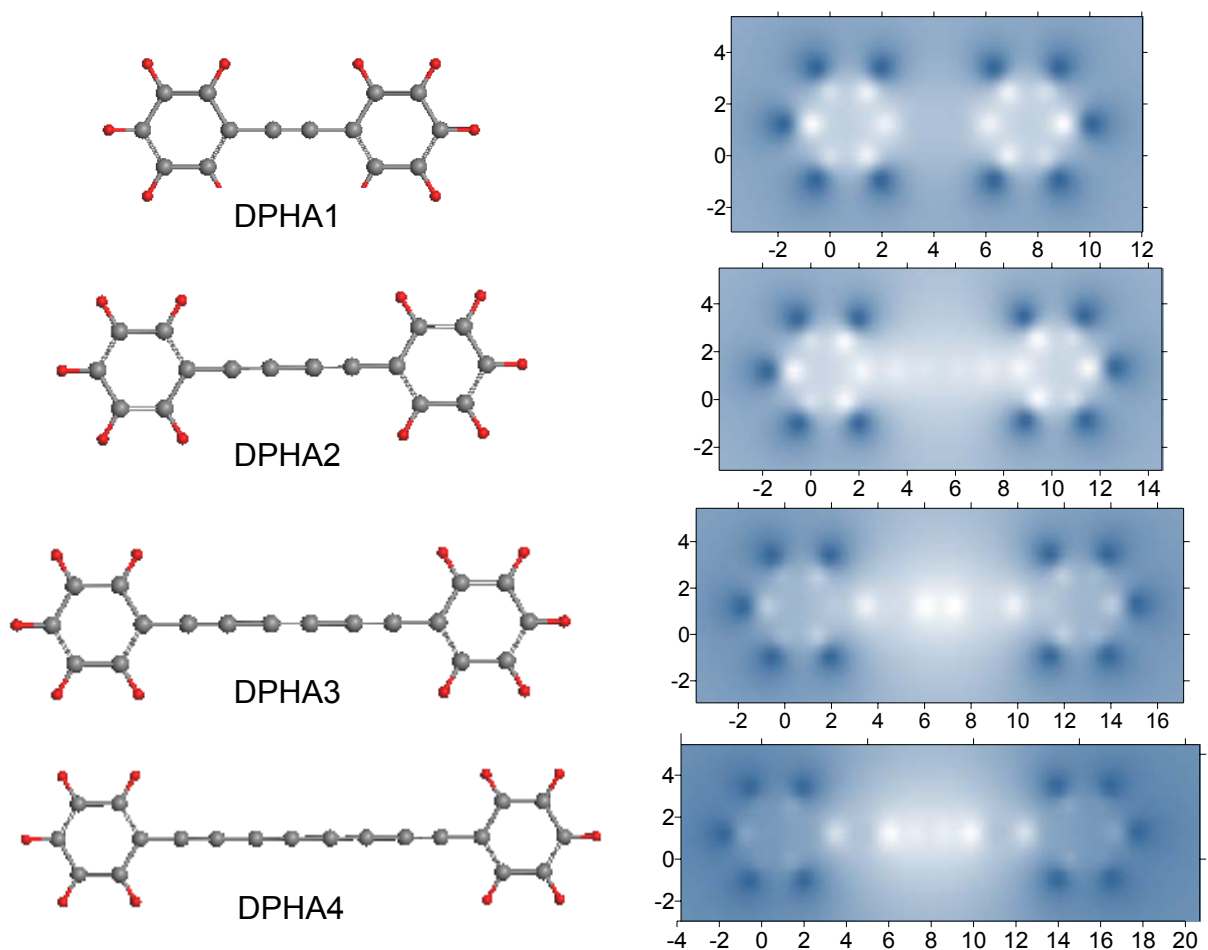

**Figure 13.** Equilibrium structures (left) and ACS maps (right) of diphenylacetylens from DPHA1 to DPHA4. The maximum value of the map intensity scale varies from 0.10$e$ for DPHA1 and DPHA2 to 0.16$e$ for DPHA3 and 0.30$e$ for DPHA4. The maps axes are in Å. Atom marking see in the caption to Fig.12. AM1-UHF calculations.

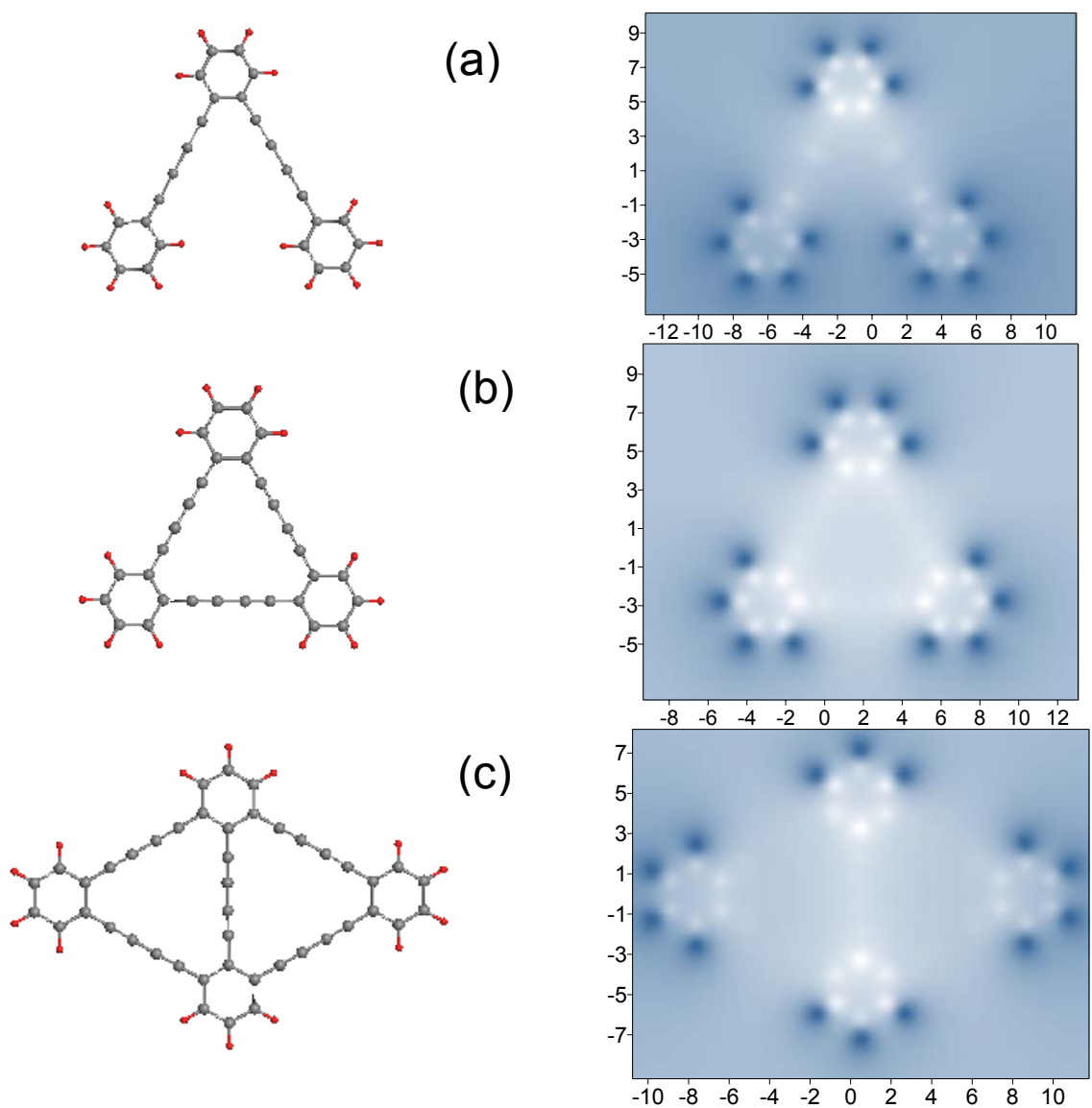

**Figure 14.** Equilibrium structures (left) and ACS maps (right) of triangle DPHA2-based compositions with one-, two-, and three-branched benzene rings (see text). The maximum value of the map intensity scale varies from 0.17$e$ in (a) and (b) to 0.23$e$ in (c). The maps axes are in Å. Atom marking see in the caption to Fig.12. AM1-UHF calculations.

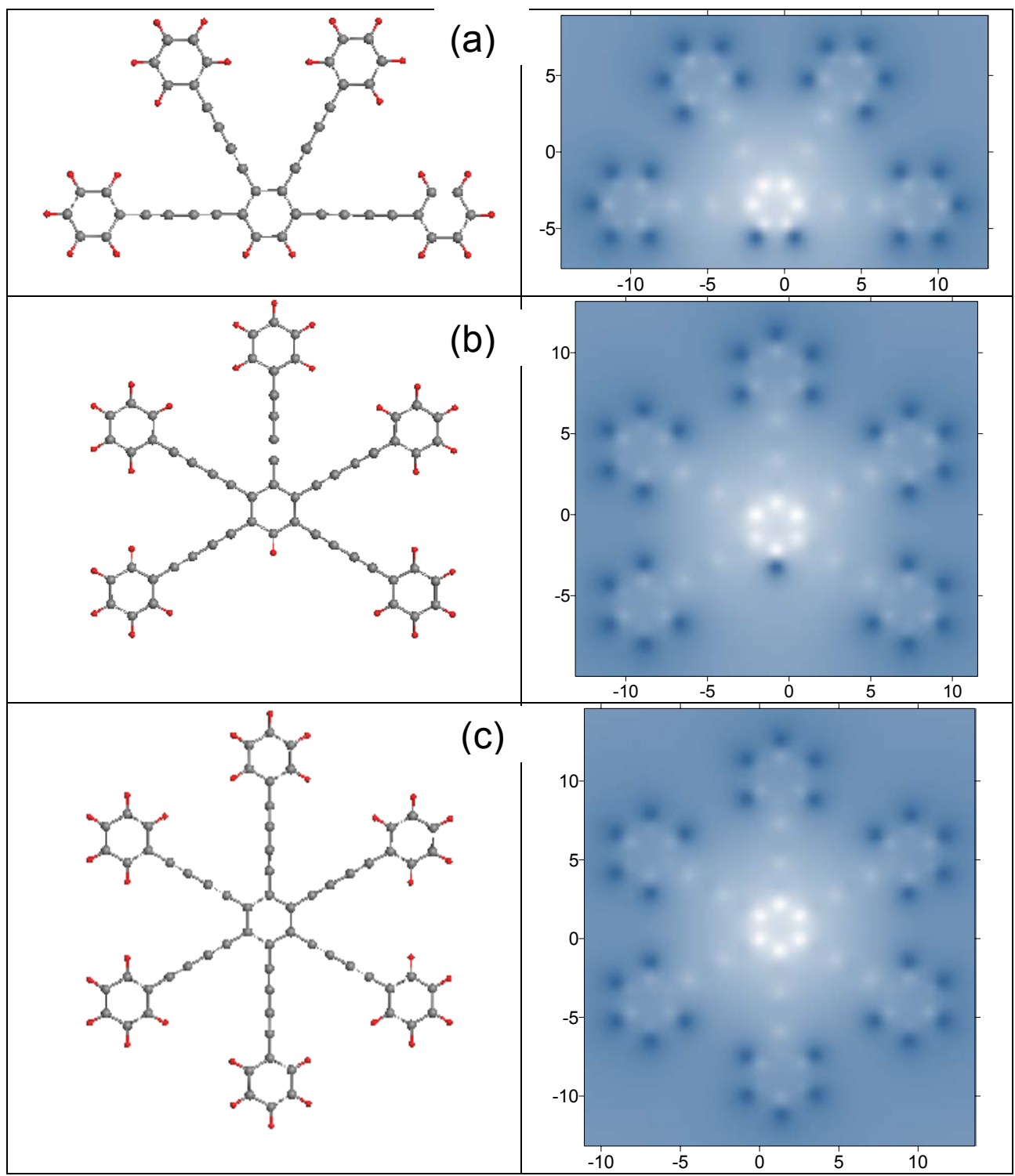

**Figure 15.** DPHA2-based GY patterns: equilibrium structures (left) and ACS (right) of four-branched (I), five-branched (II), and six-branched (III) benzenoid ring patterns (see text). The maximum value of the map intensity scale varies from 0.23$e$ in (a) to 0.25$e$ in (b), and 0.28$e$ in (c). The maps axes are in Å. Atom marking see in the caption to Fig.12. AM1-UHF calculations.

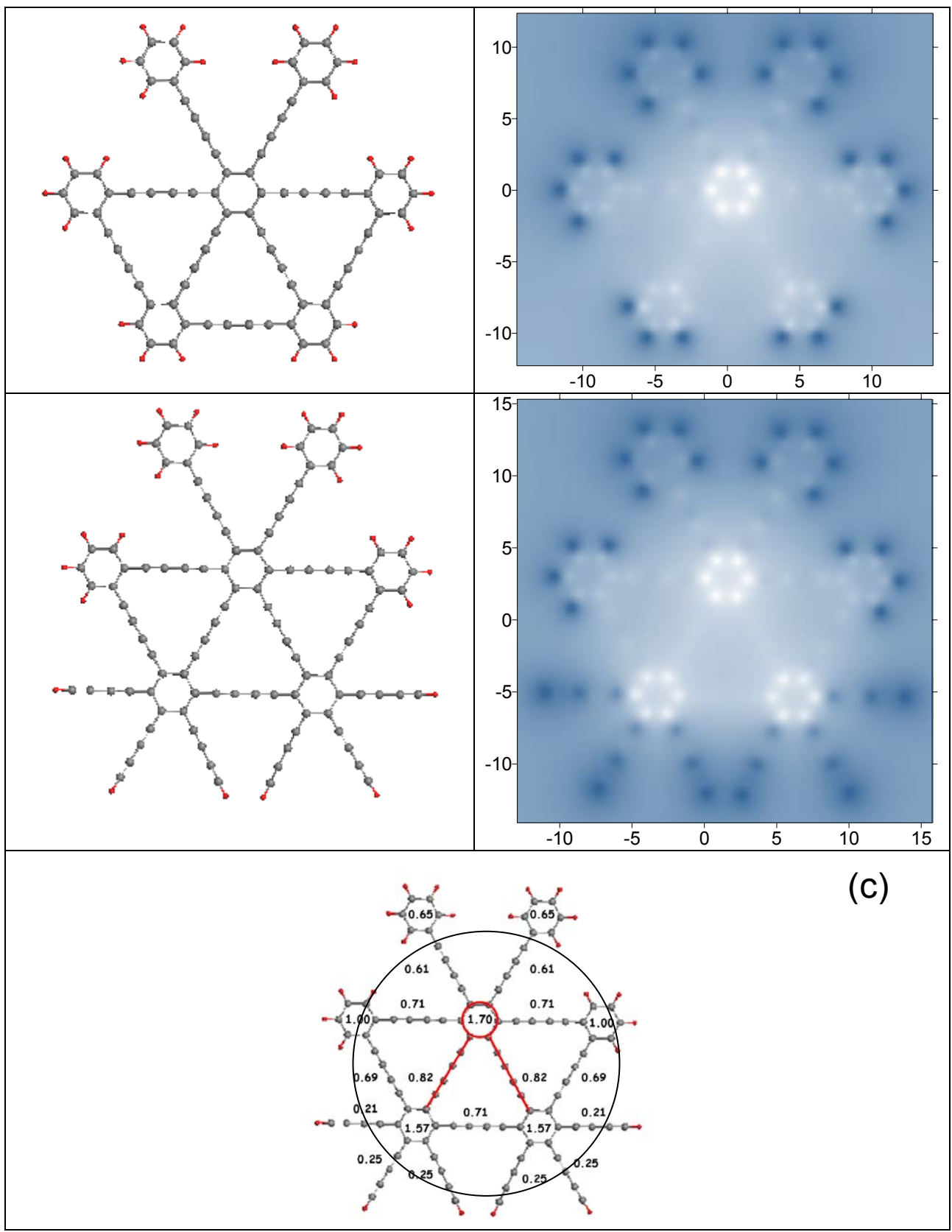

**Figure 16.** (a) and (b) Equilibrium structures (left) and ACS maps (right) of two 'Irish lace' *GDY* patterns II and III (see text). The maximum value of the map intensity scale is 0.30*e* in both cases. The maps axes are in Å. (c) The MCS map of rings and ligaments of pattern III. Large circle selects the main motif. Small circle and lines in red mark the highest MCS values characteristic for extended *GDY* structure. Atom marking see in the caption to Fig.12. AM1-UHF calculations.

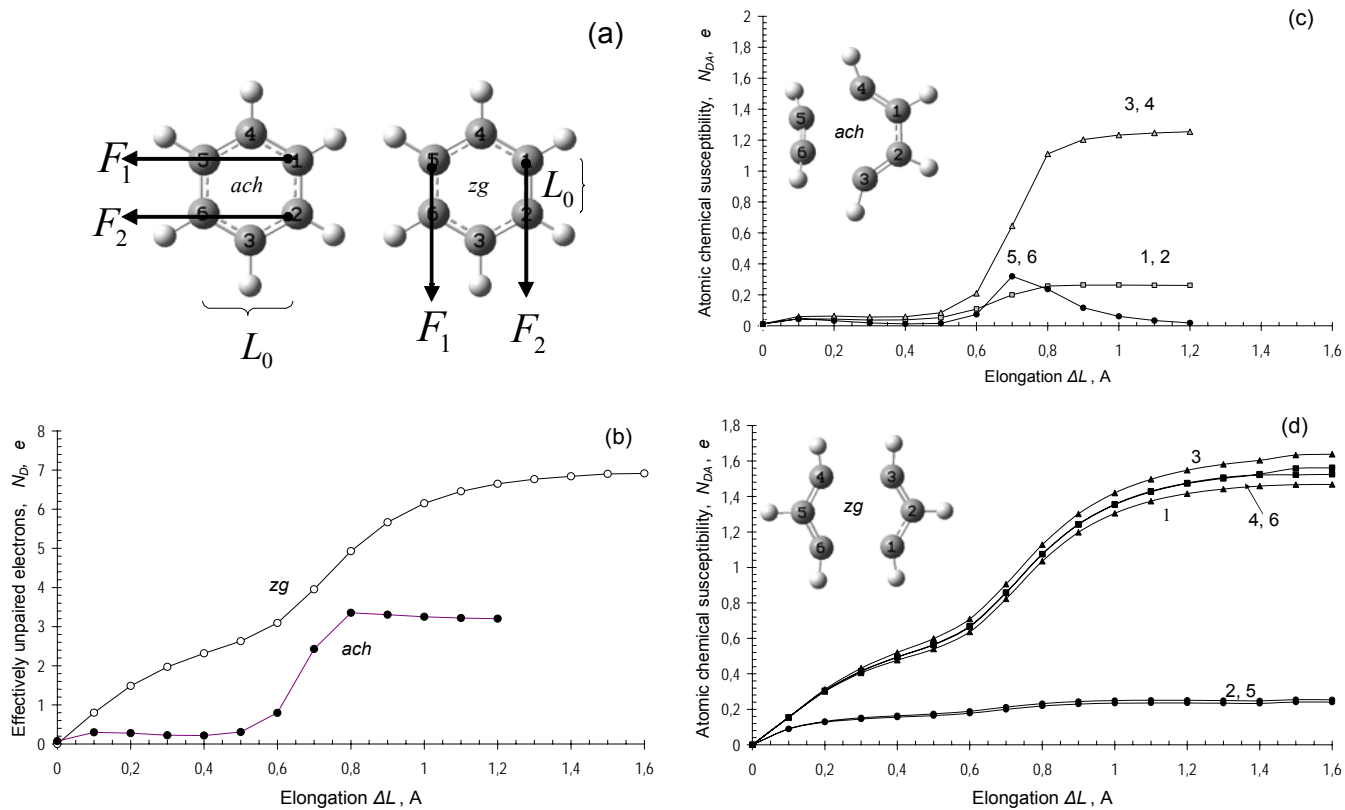

**Figure 17**. (a) Two MICs of uniaxial tension of the benzene molecule for the *ach* and *zg* deformational modes. $L_0$ indicates the initial length of the MICs while $F_1$ and $F_2$ number the corresponding forces of response. (b) $N_D(R)$ graph related to the deformation of the benzene molecule at two modes. (c) and (d) $N_{DA}(R)$ graphs related to the molecule atoms at two modes. Inserted presents products formed in due course of the benzene molecule rupture. AM1-UHF calculations.

**Table 1.** Characteristic interatomic distances related to selected covalent chemical bonds, E

| Bond | Molecule | $R_{st}^{*}$ | $R_{eq}$ | $R_{cov}$ | $W_{cov}$ | $\delta W_{cov}\%$ | $R_{k1}^{**}$ | $R_{k2}^{**}$ | $W_{rad}$ | $\delta W_{rad}\%$ |
|---|---|---|---|---|---|---|---|---|---|---|
| C-C | ethane |  | 1.503 | **2.110** | 0.607 | 40.4 |  |  |  |  |
|  | cyclohexane | 1.47-1.54 | 1.461-1.485 | 2.570 | 1.109-1.085 | 75.9-73.1 |  |  |  |  |
|  | hexamethylcyclohexane |  | 1.515-1.532 | 2.380 | 0.865-0.848 | 57.1-55.3 |  |  |  |  |
| C=C | ethylene |  | 1.326 | 1.388 | 0.062 | 4.7 | **2.140** |  | 0.752 | 56.7 |
|  | benzene | 1.34 | 1.395 | 1.395 | 0 | 0 | 2.139 |  | 0.744 | 53.3 |
|  | hexamethylbenzene |  | 1.395 | 1.408 | 0.013 | 0.9 | 2.158 |  | 0.740 | 53.0 |
| C?C | propyne | 1.20 | 1.197 | 1.240 | 0.043 | 3.6 | 1.450 | **2.100** | 0.860 | 71.8 |
| N?N | dinitrogen | 1.10 | 1.105 | 1.280 | 0.175 | 15.8 | - | 1.780 | 0.500 | 45.2 |
| C-H | ethane |  | 1.117 | 1.717 | 0.600 | 53.7 |  |  |  |  |
|  | ethylene | 1.09 | 1.098 |  |  |  |  |  |  |  |
|  | propyne |  | 1.059 |  |  |  |  |  |  |  |
| C-O | ethylene glycol | 1.43 | 1.412 | 2.000 | 0.588 | 41.6 |  |  |  |  |
| N-N | hydrazine | 1.45 | 1.378 | 1.840 | 0.462 | 33.5 |  |  |  |  |
| F-F | fluorine molecule | 1.42 | 1.427 | 1.600 | 0.173 | 12.1 |  |  |  |  |
| Si-Si | disilane | 2.35 | 2.418 | **2.950** | 0.532 | 22.0 |  |  |  |  |
| Ge-Ge | digermane | 2.44 | 2.367 | **3.000** | 0.633 | 26.7 |  |  |  |  |
| Sn-Sn | distannane | 2.81 | 2.749 | **3.550** | 0.801 | 29.1 |  |  |  |  |
| Si=Si | disilene | 2.14-2.16 | 2.293 | 1.80 |  |  | **2.800** |  |  |  |
| Ge=Ge | digermene | 2.21-2.35 | 2.324 | 2.10 |  |  | **3.000** |  |  |  |
| Sn=Sn | distannene | 2.77 | 2.161 | 2.550 | 0.389 | 18.0 | **3.600** |  | 1.439 | 66.6 |
| Si?Si | disilyne | 2.06 | 2.31 | 1.64 |  |  | - | **2.65** |  |  |
| Ge?Ge | digermyne | - | 2.30 | 1.79 |  |  | - | **2.80** |  |  |
| Sn?Sn | distannyne | 2.67 | 2.53 | 2.19 |  |  | - | **3.50** |  |  |
| Si-H | disilane [40] | 1.485 | 1.466 |  |  |  |  |  |  |  |
|  | disilene*** | 1.475-1.483 | 1.457 |  |  |  |  |  |  |  |
| Ge-H | digermane [41] | 1.541 | 1.548 |  |  |  |  |  |  |  |
|  | digermene*** | 1.530- | 1.548 |  |  |  |  |  |  |  |

**Table 2**. $X = X$ bond length, E, and molecular chemical susceptibility ($N_D$), e. in the benzene-like and (5x5) honeycomb tetrenes.

| Molecule | | C | Si | Ge | Sn |
|---|---|---|---|---|---|
| $X_6H_6$ | $R_{eq}^{db}$ | 1.395 | 2.293 | 2.026 | 2.544 (4), 2.593 (2)* <br> 2.256** |
| | $N_D$ | 0.05 | 2.68 | 0 | 1.03 <br> 0** |
| X-(5x5) | $R_{eq}^{db}$ | 1.291-1.469*** | 2.214-2.330*** | 1.941-2.407*** | 2.023-2.709*** |
| | $N_D$ | 16.63 | 42.51 | 5.56 | 10.96 |

\* The shortest and longest bonds of the molecule in Fig. 8e and d.
\*\* The data are related to the molecule in Fig. 8g.
\*\*\* The data are related to equilibrium structures in Fig. 9.

**Table 3.** Bond lengths, E, molecular ($N_D$) and atomic ($N_{DA}$) chemical susceptibilities, e, of p-dietylbenzene, diphenylacetylenes (DPHAs) and triangle-like DPHA2-based compositions (see text)[*]

| molecule | C=C bonds | C≡C bonds | C-C bond ring-alkyne and alkyne-alkyne | $N_D$ | $N_{DA}$ over ring atoms | $N_{DA}$ over triple bond atoms |
|---|---|---|---|---|---|---|
| p-debz | 1.404(4); 1.391(2); | 1.198 | 1.408(2) | 0.482 | 0.074(2); 0.068(4) | 0.021(2) |
| DPHA1 | 1.408(4); 1.392(2) | 1.198 | 1.407(2) | 1.221 | 0.101-0.082 | 0.060(2) |
|  | 1.407(4); 1.393(2) |  |  |  | 0.101-0.082 |  |
| DPHA2 | 1.410(2); 1.398(2); 1.395(2); | 1.208(2) | 1.400(2); 1.339(2)[**] | 1.515 | 0.103-0.085 | 0.092(2); 0.087(2) |
|  | 1.411(2); 1.394(2); 1.399(2) |  |  |  | 0.103-0.085 |  |
| DPHA3 | 1.411(2); 1.398(2); 1.394(2) | 1.212(2); 1.215[***] | 1.398(2); 1.339(2)[**] | 2.179 | 0.116-0.0.92; | 0.166(2)[***]; 0.154(2); 0.128(2) |
|  | 1.411(2); 1.398(2); 1.394(2) |  |  |  | 0.116-0.093 |  |
| DPHA4 | 1.412(2); 1.399(2); 1.394(2) | 1.217(2); 1.225(2)[***] | 1.395(2); 1.334(2)[**] | 3.430 | 0.132(2); 0.127; 0.122; 0.102(2) | 0.299(2)[***]; 0.269(2)[***]; 0.245(2); 0.180(2) |
|  | 1.412(2); 1.399(2); 1.394(2) |  |  |  | 0.133(2); 0.127; 0.122; 0.102(2) |  |
| composition 1 (Fig. 14a) | 1.412-1.394 | 1.211(2); 1.209(2) | 1.399(2); 1.395(2); 1.343(2)[**] | 3.041 | 0.162(2); 0.146(2); 0.142(2) **0.90**[****] | 0.121(2); 0.119(2); 0.115(2); 0.102(2) |
|  | 1.412-1.394 |  |  |  | 0.109(2); 0.108; 0.104; 0.089(2) **0.61** | **0.455**(2) per a ligament |
|  | 1.412-1.394 |  |  |  | 0.109(2); 0.108; 0.104; 0.089(2) **0.61** |  |
| composition 2 (Fig. 14b) | 1.412-1.394 | 1.212(6) | 1.395(6); 1.341(3)[**] | 4.536 | 0.170(2); 0.154(2); 0.149(2) **0.95** | 0.144(4); 0.138(4); 0.143(2); 0.137(2) |
|  | 1.411-1.394 |  |  |  | 0.170(2); 0.154(2); | **0.56**(3) per a ligament |
| composition 3 (Fig. 14c) | 1.411-1.394 | 1.212(8); 1.214(2) | 1.395(8); 1.391(2); 1.340(4)[**]; 1.338[**] | 7.497 | 0.170(2); 0.154(2); 0.149(2) **0.95** | 0.155(4); 0.153(4); 0.152(4); 0.144(4); 0.179(2); 0,174(2) |
|  | 1.412-1.394 |  |  |  | 0.170(2); 0.154(2); 0.149(2) **0.95** |  |
|  | 1.428(2); 1.411(2); 1.396(2) |  |  |  | 0.225; 0.207(2); 0.201(2); 0.175 **1.22** |  |
|  | 1.412(2); 1.395(2); 1.394(2) |  |  |  | 0.173(2); 0.157(2); 0.151(2) **0.96** | **0.71**(1) and **0.60**(4) per a ligament |
|  | 1.412(2); 1.395(2); 1.394(2) |  |  |  | 0.173(2); 0.157(2); 0.151(2) **0.96** |  |
|  | 1.428(2); 1.411(2); 1.396(2) |  |  |  | 0.225; 0.207(2); 0.201(2); 0.175 **1.22** |  |

[*] Figures in parentheses number identical structural units.
[**] The distance between two alkyne units.
[***] Inner alkyne unit.
[****] Bold numbers correspond to the MCSs of individual rings and ligaments.

**Table 4.** Bond lengths, $E$, molecular ($N_D$) and atomic ($N_{DA}$) chemical susceptibilities, $e$, of DPHA-based 'Irish lace' patterns (see text)[*]

| pattern | C=C bonds | C?C bonds | C-C bonds ring-alkyne and alkyne-alkyne | $N_D$ | $N_{DA}$ and $N_D$ (bold) over rings | $N_D$ over all and individual (bold) ligaments |
|---|---|---|---|---|---|---|
| pattern I (Fig. 15a) | 1.426(2); 1.412(2); 1.390(2) <br> 1.411(2); 1.399(2); 1.394(2) | 1.212-1.210 | 1.398-1.394; 1.341[**] | 6.119 | 0.235(2); 0.230(2); 0.210(2) <br> **1.35** <br> 0.116; 0.115; 0.113; 0.110; 0.092(2) <br> **0.64** | 2.20 <br> **0.55** (4) |
| pattern II (Fig. 15b) | 1.428(2); 1.426(2); 1.408(2) <br> 1.411(2); 1.399(2); 1.394(2) | 1.213-1.211 | 1.399-1.392; 1.342[**] | 7.615 | 0.265(2); 0.257; 0.250; 0.236(2) <br> **1.51** <br> 0.117(2); 0.115; 0.110; 0.093(2) <br> **0.64** | 2.90 <br> **0.58** (5) |
| pattern III (Fig. 16a) | 1.426(6) <br> 1.411(2); 1.398(2); 1.394(2) | 1.214-1.211 | 1.392; 1.341[**] | 9.203 | 0.274(6) <br> **1.64** <br> 0.118(2); 0.115; 0.111; 0.094(2) <br> **0.65** | 3.66 <br> **0.61** (6) |
| pattern IV (Fig. 16b) | 1.427(2); 1.426(4) <br> 1.428(2); 1.411(2); 1.391(2) <br> 1.412(2); 1.402(2); 1.394(2) <br> 1.411(2); 1.398(2); 1.394(2) | 1.214-1.211 | 1.392; 1.341[**] | 13.489 | 0.281(6) <br> **1.69** <br> 0.227; 0.210(2); 0.203(2); 0.177 <br> **1.23** <br> 0.177-0.153 <br> **0.98** <br> 0.118(2); 0.111(2); 0.094(2) <br> **0.65** | 6.08 <br> **0.77** (2); **0.70** (5); **0.57**(2) |
| pattern V (Fig. 16c) | 1.427(2); 1.426(2); 1.425(2) <br> 1.428(2); 1.411(2); 1.391(2) <br> 1.412(2); 1.402(2); 1.394(2) <br> 1.411(2); 1.398(2); 1.394(2) | 1.214-1.211 | 1.392; 1.341[**] | 16.144 | 0.282-0.284 <br> **1.70** <br> 0.271-0.251 <br> **1.57** <br> 0.177-0.153 <br> **0.98** <br> 0.118(2); 0.111(2); 0.094(2) <br> **0.65** | 6.08 <br> **0.82** (2); **0.71** (3); **0.69** (2); **0.61** (2); **0.25** (4); **0.21** (2) |

[*] Figures in parentheses number identical chemical bonds and ligaments.
[**] The distance between two alkyne units.